\documentclass[11pt]{article}

\usepackage{amsmath,amssymb,amsthm,color,epsfig}
\usepackage{amsbsy}

\usepackage{epsfig}
\usepackage{mathrsfs}

\usepackage{amsthm}
\usepackage{amsmath,amssymb,amsbsy,color}

\newtheorem{theorem}{Theorem}

\newtheorem{definition}[theorem]{Definition}

\newtheorem{lemma}[theorem]{Lemma}

\newtheorem{problem}[theorem]{Problem}

\newcommand{\col}[3]{ \renewcommand{\arraystretch}{#1}
                \left[\!\! \begin{array}{c} #2 \\ #3 \end{array} \!\!\right] }

\newcounter{geqncount}
    {\refstepcounter{equation}%
     \setcounter{geqncount}{\value{equation}}%
     \setcounter{equation}{0}%
  }%
    {\setcounter{equation}{\value{geqncount}}}

\newcommand{\half}{{\text{$\textstyle\frac{1}{2}$}}}

\newcommand{\Z}{\mathbb{Z}}
\newcommand{\R}{\mathbb{R}}
\newcommand{\C}{\mathbb{C}}
\newcommand{\M}{\mathbb{M}}
\newcommand\sysone{(H_1,\Omega_1)}
\newcommand\systwo{(H_2,\Omega_2)}

\newcommand\sysbig{({\cal H},\Omega)} 

\newcommand{\QED}{\rule{0.4em}{2ex}}
\newcommand{\im}{\mathrm{ Im}}
\newcommand{\re}{\mathrm{ Re}}
\newcommand{\wk}{\widetilde{\kappa}}
\newcommand{\wo}{\widetilde{\omega}}
\newcommand{\bigo}{{O}}
\newcommand{\FT}{{\cal F}}
\newcommand{\thph}{(\theta,\phi)}
\newcommand{\kw}{(\kappa,\omega)}
\newcommand{\kwz}{(\kappa_0,\omega_0)}

\newcommand\slantfrac[2]{\hbox{$\,^{#1\hspace{-1pt}}\!/_{\hspace{-1pt}#2}$}}
\newcommand\onehalf{\slantfrac{1}{2}}

\begin{document}

\bibliographystyle{plain} 

\begin{center}
{\bf \Large  A lattice model for}
{\bf \Large resonance in open periodic waveguides}
\end{center}

\vspace{0.2ex}

\begin{center}
{\scshape \large Natalia Ptitsyna, Stephen P. Shipman}
\end{center}

\begin{center}
{\itshape Department of Mathematics, Louisiana State University} \\
{\itshape Baton Rouge, LA  \ 70803, \ USA} 
\end{center}

\vspace{3ex}
\centerline{\parbox{0.9\textwidth}{
{\bf Abstract.}\ We present a discrete model of resonant scattering of waves by an open periodic waveguide.  The model elucidates a phenomenon common in electromagnetics, in which the interaction of plane waves with embedded guided modes of the waveguide causes sharp transmission anomalies and field amplification.  The ambient space is modeled by a planar lattice and the waveguide by a linear periodic lattice coupled to the planar one along a line.  We show the existence of standing and traveling guided modes and analyze a tangent bifurcation, in which resonance is initiated at a critical coupling strength where a guided mode appears, beginning with a single standing wave and splitting into a pair of waves traveling in opposing directions.
Complex perturbation analysis of the scattering problem in the complex frequency and wavenumber domain reveals the complex structure of the transmission coefficient at resonance.
}}

\vspace{3ex}
\noindent
\begin{mbox}
\end{mbox}

\vspace{3ex}
\hrule
\vspace{2ex}

\section{Motivation}

When a periodic waveguide is in contact with an ambient space, the interaction between modes of the guide and radiation originating from the ambient space outside the guide results in interesting resonant behavior.  The resonance is manifest by pronounced amplitude enhancement of fields in the waveguide and sharp anomalies in the graph of transmitted energy versus frequency near the frequency of the guided mode \cite[{\itshape e.g.}\hspace{0ex}]{FanJoannopoul2002,KanskarPaddonPacradouni1997,LiuLalanne2008}.  Examples abound in the physics and engineering literature because of the importance of these anomalies in applications to photo-electronic devices.  Because of the exchange of energy between the waveguide and the surrounding space, we call the waveguide {\itshape open}.

In this work, we analyze a discrete model of resonance in open periodic waveguides.  The ambient space is modeled by a uniform two-dimensional lattice, and the waveguide is modeled by a periodic one-dimensional lattice, coupled to the two-dimensional one along a line.  With period two, we consider this to be the simplest model that exhibits the essential features of the open lossless waveguide in air.  It is not unlike the Anderson model, in which a single chain of beads interacts with a resonator attached to one of the beads \cite{Longhi2007,Mahan1993,MiroshnicMingaleevFlach2005}.  But, unlike the Anderson model, a periodic model, with period at least~2, admits both propagating and evanescent Fourier harmonics simultaneously, and this is precisely the feature of open periodic waveguides that allows embedded guided modes and their resonant interaction with incident radiation.

The discrete model is useful in that it exhibits important resonant phenomena of continuous open waveguides while its simplicity permits explicit calculations and proofs.  In particular, one can prove that the resonant peaks and dips in the transmitted energy reach exactly 100\% and 0\%, a phenomenon that is often observed in scattering of electromagnetic waves by open waveguides.  Moreover, explicit formulas illuminate the connection between structural parameters of the waveguide and the properties of the anomalies, such as central frequency and width, both of which are important in applications of lasers and LEDs \cite{FanVilleneuveJoannopoul2000}.

In addition, we analyze a tangent bifurcation of resonances, in which resonance is initiated at a critical coupling strength, beginning with a single standing wave and splitting into a pair of waves traveling in opposing directions.

\smallskip

The kind of resonance we are describing here is akin to those that go by the names of Feshbach resonance, Breit-Wigner resonance, or Fano resonance in quantum mechanics.  The unifying idea is that, when one perturbs a system that admits a bound state whose frequency is embedded in the continuous spectrum, the eigenvalue dissolves as a result of the coupling of the bound state to the extended states corresponding to the frequencies of the continuum.  This coupling is the cause of sharp features in observed scattering data near the bound state frequency \cite[\S XXII.6]{ReedSimon1980d}, \cite{Fano1961,WeisskopfWigner1930}.

A similar type of resonance of classical fields (as electromagnetic or acoustic) results from the interaction of guided modes of a periodic waveguide and plane waves originating from outside the guide.  For this interaction to take place, the waveguide must be open, that is, it must be in contact with the ambient space, such as in the case of a photonic crystal slab in air.  Because of the periodicity of the waveguide, guided modes may couple to the Rayleigh-Bloch diffracted waves and become ``leaky", or ``quasiguided" modes, or ``guided resonances" \cite{FanJoannopoul2002,PengTamirBertoni1975,TikhodeevYablonskiMuljarov2002}.  These leaky modes are associated with the sharp anomalies in the graph of the transmitted energy across the slab as a function of frequency, as we have mentioned.

Under certain conditions, a lossless open periodic waveguide can actually support a true guided mode---one that is exponentially confined to the guide.  This can occur in one of two ways: (1) if, for a given frequency and Bloch wavevector parallel to the guide, the expansion of the fields in spatial Fourier harmonics parallel to the guide admits no harmonics that propagate away from the guide (this is the region below the light cone in the first Brillouin zone); or (2) there are Fourier harmonics that propagate away from the waveguide (the Rayleigh-Bloch diffracted waves, or propagating diffractive orders) but the structure admits a field for which the coefficients of these harmonics {\em happen to vanish}.  The latter occurs, for example, in a waveguide that is symmetric about a plane perpendicular to it at a wavevector parallel to the plane of symmetry \cite{Bonnet-BeStarling1994,ShipmanVolkov2007,TikhodeevYablonskiMuljarov2002}.

We will be concerned with the latter type of guided mode, for only these can interact with incident radiation.  Such modes are typically nonrobust and are therefore excited by small deviations of the angle of incidence (associated with the Bloch wavevector) or perturbations of the structure.  Their frequencies can be viewed as embedded eigenvalues in a pseudoperiodic scattering problem for a fixed Bloch wavevector.  Perturbation of the wavevector or the structure itself destroys the guided mode and the associated eigenvalue.  In 
\cite{ShipmanVenakides2005}, Shipman and Venakides derived, for the two-dimensional case, an asymptotic formula for the transmitted field as a function of wavevector and frequency based on complex perturbation analysis of the scattering problem about the guided mode parameters.  The formula is rigorous and subsumes that derived by Fano in the context of quantum mechanics \cite{Fano1961}; it plays a major role in the analysis of resonance in this paper.  An in-depth discussion of resonance near nonrobust guided modes can be found in~\cite{Shipman2010}, and the role of structural asymmetry on the detuning of resonance is analyzed in a discrete model in \cite{ShipmanRibbeckSmith2010}.

\medskip

The exposition proceeds as follows.

{\bfseries \S 2. {\itshape The discrete model.}} First, we describe the ambient and waveguide components of the discrete model and how they are coupled.  We then identify the minimal space of motions in the ambient 2D lattice (the ``reconstructible" part), which, together with the 1D waveguide, form a closed system, decoupled from a complementary ``frozen" space of motions of the 2D lattice.

{\bf \S 3. {\itshape The scattering problem.}}  The problem of scattering of plane waves in the 2D lattice by the 1D waveguide is posed.  We describe the Fourier decomposition of fields and resolution of scattered fields into their diffractive orders and prove existence of solutions, including in the presence of a guided mode.

{\bf \S 4. {\itshape Guided modes.}}  We prove the existence or nonexistence of true guided modes under certain conditions and describe the dispersion relation for generalized (leaky) modes relating complex frequency to complex wavenumber.  A nonrobust embedded guided mode is characterized by an isolated point in the real frequency-wavenumber plane that lies on the complex dispersion relation.

{\bf \S 5. {\itshape Resonant scattering}}  near a guided-mode frequency.  This is the most important and interesting section of the paper.  The complex perturbation analysis of \cite{ShipmanVenakides2005} is applied to transmission anomalies for the discrete model.  We extend it to capture the singular behavior of the transmitted energy at a tangent bifurcation of guided modes (Theorem~\ref{thm:bifurcation}), in which the bifurcation parameter is a constant of coupling between the ambient lattice and the waveguide.  We also analyze the accompanying resonant amplification.


\section{The Discrete Model}

We have chosen to analyze a model in which the ambient space and the wave guide can first be described as separate systems in their own right, which are then coupled together through simple coupling constants along a line.


\subsection{The Ambient Planar Lattice}

The ambient space is a planar (two-dimensional) lattice of beads of mass $1$ located at the integer points $\Z^2$ in $\R^2$ that are connected by springs of strength $1$. The internal dynamics are given by a
Schr\"odinger-type equation
\begin{equation}
\label{first}
\dot y = -i \Omega_2 y,
\end{equation}  
where $y=\{y_{mn}\} \in \ell^2(\mathbb{Z}^2)=:H_2$ with $(m,n) \in \Z^2$ and $-\Omega_2$ is the discrete uniform Laplacian
\begin{equation}
\label {second}
(\Omega_{2} y)_{mn} =-(y_{(m-1)n}+y_{(m+1)n}+y_{m(n-1)}+y_{m(n+1)}-4y_{mn}).
\end{equation}
The spatial part $u = \{u_{mn}\}$ of a harmonic solution $y_{mn}(t)=e^{-i\omega t}u_{mn}$ satisfies
\begin{equation}
\label{third}
(\Omega_2-\omega)u_{mn}=0.
\end{equation}
The solutions of this equation are generalized eigenfunctions of $\Omega_2$, and the simplest of these are the plane waves
\begin{equation}\label{dispersionrelation}
  e^{2 \pi i (m \theta +n \phi)}, \quad \omega = 4 - 2\cos(2\pi\theta) - 2\cos(2\pi\phi).
\end{equation}
This relation between $\omega$ and $\thph$ is the dispersion relation for the free 2D lattice.

Through the (inverse) Fourier transform, each element of $H_2$ is expressed as an integral superposition of these eigenfunctions
\begin{equation*}
\label{fourth}
  \FT: H_2 \to L^{2}([-\onehalf,\onehalf]^2), \quad f(\theta,\phi) = (\FT u)\thph = \sum_{m=-\infty}^{\infty}
  \sum_{n=-\infty}^{\infty}u_{mn}e^{-2 \pi i (m \theta +n \phi)},
\end{equation*}
\begin{equation*}
  u_{mn} = (\FT^{-1} f)_{mn} = \int_{-\half}^\half\int_{-\half}^\half f\thph e^{2 \pi i (m \theta +n \phi)} d\theta\,d\phi.
\end{equation*}
The operator $\FT$ is unitary.  The bounded operator $\Omega_2$ can be written in terms of the shift operators on $H_2$,
\begin{equation*}
  \renewcommand{\arraystretch}{1.2}
\left.
  \begin{array}{ll}
    (S^{\text{r}}u)_{mn}= u_{(m-1)n}, & (S^{\text{l}}u)_{mn}= u_{(m+1)n}, \\
    (S^{\text{u}}u)_{mn}= u_{m(n-1)}, & (S^{\text{d}}u)_{mn}= u_{m(n+1)},
  \end{array}
\right.
\end{equation*}
\begin{equation}
\label{tenth}
\Omega_2=-(S^{\text{l}}+S^{\text{r}}+S^{\text{u}}+S^{\text{d}}-4\mathrm{I}).
\end{equation}
The shift operators become multiplication operators under the Fourier transform, 
\begin{equation}
\label{eleventh}
\begin{cases}
\FT S^{\text{r}} \FT^{-1} f(\theta,\phi)=e^{-2 \pi i \theta}f(\theta, \phi), \\
\FT S^{\text{l}} \FT^{-1} f(\theta,\phi)=e^{2 \pi i \theta}f(\theta, \phi), \\ 
\FT S^{\text{u}} \FT^{-1} f(\theta,\phi)=e^{-2 \pi i \phi}f(\theta, \phi), \\
\FT S^{\text{d}} \FT^{-1} f(\theta,\phi)=e^{2 \pi i \phi}f(\theta, \phi),
\end{cases}
\end{equation}
and we thus obtain a spectral representation of $\Omega_2$, in which the value of the multiplication operator at $(\theta,\phi)$ is the frequency given by the dispersion relation for plane waves above \eqref{dispersionrelation}.
\begin{equation}
(\FT \Omega_2 \FT^{-1} )f(\theta, \phi)
= (4-2\cos( 2 \pi \theta )-2 \cos ( 2 \pi \phi )) f(\theta, \phi).
\end{equation}
The spectrum of $\Omega_2$ is the range of this multiplier, $[0,8]$.

\subsection{The Periodic Waveguide}

Our periodic waveguide is an infinite sequence of beads connected by springs.  The internal dynamics in the Hilbert space $H_1 = \ell^2(\Z)$ are given by the equation
\begin{equation}
\label{thirteen}
\M\dot x = -i A x,
\end{equation}
where $\M$ is the positive mass operator defined by
\begin{equation}
\label{fourteenth}
(\M x)_j := M_j x_j, \quad M_j>0,
\end{equation}
the internal operator $A$ is minus the discrete nonuniform Laplacian
\begin{equation}
\label{fifteenth}
(Ax)_j := -k_j x_{j+1}+(k_j+k_{j-1}) x_j - k_{j-1}x_{j-1},
\end{equation}
and both $\M$ and $A$ are taken to be $N$-periodic:
\begin{equation}
\label{sixteenth}
M_{j+N} = M_j \text{ and } k_{j+N} = k_j \quad \text{for all } j\in\Z.
\end{equation}
By  redefining the variable by the substitution $x\mapsto\M^{-\frac{1}{2}}x$ and denoting the operator $\M^{-\frac{1}{2}} A \M^{-\frac{1}{2}}$ by $\Omega_1$, we reduce the equation to a simpler form:
\begin{equation}
\label{eighteenth}
\dot x = -i\Omega_1 x.
\end{equation}
Since $A$ and $\M$ are self-adjoint, so is $\Omega_1$, and it is represented by a tridiagonal matrix with periodic entries,
\begin{equation}
\label{nineteenth}
(\Omega_1 x)_j= -\frac{k_j}{\sqrt{M_j M_{j+1}}}x_{j+1}+\frac{(k_j +k_{j-1})}{M_j}x_j-\frac{k_{j-1}}{\sqrt{M_j M_{j-1}}}x_{j-1}.
\end{equation}
Since $\Omega_1$ commutes with the shift operator $S$,
\begin{equation}
\label{twentieth}
(Sx)_j=x_{j+N},
\end{equation}
we can obtain by the Floquet theory the generalized eigenfunctions of $\Omega_1$ by examining those of~$S$.  Since $S$ is unitary, its generalized eigenfunctions $\{x_j\}$ are characterized by the pseudo-periodic condition
\begin{equation}
\label{twenty-second}
x_{j+N} = e^{2 \pi i\kappa}x_j,  \quad  -\half < \kappa \leq \half.
\end{equation} 
Let us denote by $\mathcal{P}_{\kappa}$ the $N$-dimensional space of solutions, which is spanned by the vectors ($\ell=1,\ldots,N$): 
\begin{equation}
\mathfrak{p}^{(\ell)}=(\ldots,\underbrace{e^{-2 \pi i \kappa}}_{-N+\ell}
,0,\ldots, 0, \underbrace{1}_{\ell}
, 0, \ldots,0,\underbrace{e^{ 2 \pi i \kappa}}_{N+\ell}
,0 ,\ldots, 0,\underbrace{e^{ 4 \pi i \kappa}}_{2N+\ell}
,\ldots) . 
\end{equation}
%
%
%
%
With respect to this basis, the restriction $\Omega_1^{(\kappa)}$ of $\Omega_1$ to $\mathcal{P}_{\kappa}$ is represented by the Floquet matrix
\begin{equation}
\label{twenty-fourth}
\Omega_{1}^{(\kappa)}=
\begin{pmatrix}
\!\frac{(k_1+k_N)}{M_1}\! &\! \frac{-k_1}{\sqrt{M_2 M_1}}\! & \!0 \!& \!\cdots \!&\! 0\! &\! \frac{-k_N}{\sqrt{M_N M_1}} e^{2 \pi i \kappa}\! \\
\!\frac{-k_1}{\sqrt{M_2 M_1}}\! & \!\frac{(k_2+k_1)}{M_2} \!&\! \frac{-k_2}{\sqrt{M_2 M_3}}\! & \!\cdots \!&\! 0 \!& \!0\! \\
\!\vdots \!&\! \vdots\! &\! \vdots\! &\! \ddots \!&\! \vdots \!&\! \vdots\! \\
\!\frac{-k_N}{\sqrt{M_N M_1}}e^{-2 \pi i\kappa}\! &\! 0\! &\! 0 \!&\! \cdots\! &\! \frac{-k_{N-1}}{\sqrt{M_N M_{N-1}}}\! & \!\frac{(k_N+k_{N-1})}{M_N}\!
\end{pmatrix} .
\end{equation}

\begin{figure}[ht]
	\centering 
\includegraphics[width=13.8cm,height=7.0cm]{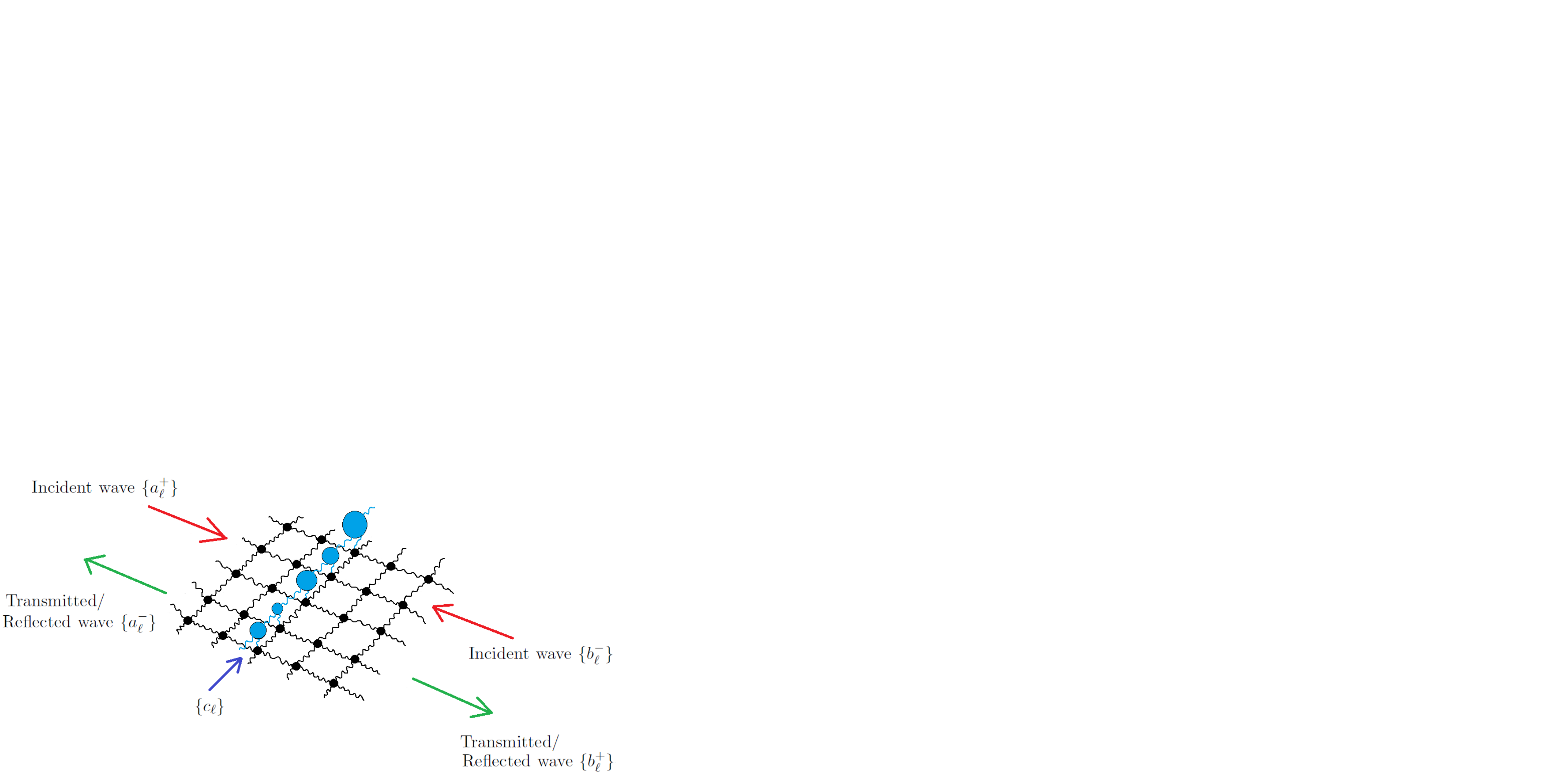}
\caption{\footnotesize{A schematic diagram of the discrete model.}}
 \label{Incident_Transmitted}
\end{figure}

\subsection{The Coupled System}

Let us couple the systems $\sysone$ and $\systwo$ in a simple way by introducing a periodic sequence of constants $\gamma_n$ with $\gamma_{n+N}=\gamma_n$ that couple $x_n$ to $u_{0n}$.  This is achieved by the coupling operator $\Gamma: H_2 \rightarrow H_1$ defined through
$$
\Gamma(E_{0n})= \gamma_n e_{n}, \quad \Gamma(E_{mn})=0, \text{ if $m \ne 0$},
$$
in which $\{e_n\}_{n\in\Z}$ and $\{E_{mn}\}_{m,n\in\Z}$ are the standard orthonormal Hilbert-space bases for $H_1$ and $H_2$, respectively.
The adjoint $\Gamma^\dagger:H_1\to H_2$ of $\Gamma$ is
$$
\Gamma^{\dag}(e_n)=\bar{\gamma}_n E_{0n}.
$$
The internal dynamics in $H_1$ and $H_2$, together with the coupling between them, define a lossless oscillatory dynamical system $\sysbig$ in the Hilbert space
\begin{equation}
\label{fifty-seventh}
{\cal H} = H_1\oplus H_2,
\end{equation}
where $\Omega$ has the following form with respect to this decomposition
\begin{equation}
\label{fifty-eighth}
\Omega = 
\left[
\begin{array}{cc}
\Omega_1 & \Gamma \\
\Gamma^\dagger & \Omega_2
\end{array}
\right].
\end{equation}
and the dynamics in $\cal H$ are given by
\begin{equation*}
  \col{1}{\dot x}{\dot y} = -i\, \Omega \col{1}{x}{y}.
\end{equation*}
The assumption of a harmonic field with circular frequency $\omega$, $x=e^{-i\omega t}z$ and $y=e^{-i\omega t}u$, leads to the eigenvalue problem
\begin{equation}
\Omega  \col{1}{z}{u} = \omega \col{1}{z}{u}. 
\end{equation}
which is equivalent to the coupled system
\begin{eqnarray}
\label{eq1}
&&\omega z_n=(\Omega_1 z)_n + (\Gamma u)_{n}\,, \\
\label{eq2}
&&\omega u_{mn}=(\Gamma^{\dagger}z)_{mn}+(\Omega_2 u)_{mn}\,.
\end{eqnarray}

Because of the periodicity of the waveguide, the operator $\Omega$ commutes with translation by $N$ lattice points in the $n$ variable, that is, $u_{mn}\mapsto u_{m,n-N}$.  By the Floquet-Bloch theory, $\Omega$ is a direct integral of pseudo-periodic operators $\Omega_\kappa$,
\begin{equation*}
  \Omega = \int_{[-\half,\half]}^\oplus \Omega_\kappa\,d\kappa,
\end{equation*}
which are defined by the restriction of $\Omega$ to the functions $u_{mn}$ that satisfy the pseudo-periodic condition
\begin{equation*}
  u_{m,n+N} = e^{2\pi i\kappa} u_{mn}.
\end{equation*}

\subsection{Dynamics projected onto the waveguide}

If the conservative system $\sysbig$ is projected onto $H_1$, the result is the dissipative system
\begin{equation}\label{dissipative}
  \dot z = -i\Omega_1 z - \int_0^\infty \Gamma e^{-i\tau\Omega_2}\Gamma^\dagger z(t-\tau)d\tau\,,
  \quad z(t) \in H_1.
\end{equation}
One can ask the question: how much of the original system $\sysbig$ can be reconstructed from the dynamics projected onto $H_1$, that is, from equation \eqref{dissipative} alone; more viscerally: which motions of the ambient lattice can be detected by an observer living in the waveguide?  Equivalently, one could ask which motions originating in the ambient lattice can disturb the waveguide.  Figotin and Schenker \cite{FigotinSchenker2005} prove that a dissipative system of the form \eqref{dissipative} admits a conservative extension $(\tilde{\cal H},\tilde \Omega)$, with dynamics $\dot{\tilde v} = -i\tilde\Omega \tilde v$ ($\tilde v(t)\in\tilde H$), that is unique up to Hilbert-space isomorphism.  The unique extension of the projection of $\sysbig$ onto $H_1$ can be realized as a unique subsystem of the original system $\sysbig$.
The construction of this subsystem is given by Theorem 9 in \cite{FigotinShipman2006},
\begin{eqnarray*}
  && \tilde{\cal H} = H_1 \oplus \tilde H_2, \quad \tilde H_2 =\mathcal{O}_{\Omega_2}(\Gamma^\dagger(H_1)), \\
  && \tilde\Omega = \Omega_1 \oplus \Omega_2|_{\tilde H_2},
\end{eqnarray*}
in which the subspace $\tilde H_2$ of $H_2$ is the orbit of the image $\Gamma^\dagger(H_1)$ in $H_2$ under the action of $\Omega_2$.  The orbit is defined by


\begin{definition}[orbit]
Let $\Omega$ be a self-adjoint operator in a Hilbert space $H$ and $S$ a subset of vectors in $H$. Then we define the closed orbit (or simply orbit) $\mathcal{O}_{\Omega}(S)$ of $S$ under action of $\Omega$ by
\begin{equation}
\mathcal{O}_{\Omega}(S)=\text{closure of span } \{\psi(\Omega)w: \psi\in C_{c}(\mathbb{R}), w \in S \},
\end{equation}
where $C_{c}(\mathbb{R})$ is the space of continuous complex-valued functions on $\R$ with compact support.
If $H'$ is a subspace of $H$ such that $\mathcal{O}_{\Omega}(H')=H'$, then $H'$ is said to be invariant with respect to $\Omega$ or simply $\Omega$-invariant.
\end{definition}

\begin{figure}[ht]
	\centering 
\includegraphics[width=5.0cm,height=4.5cm]{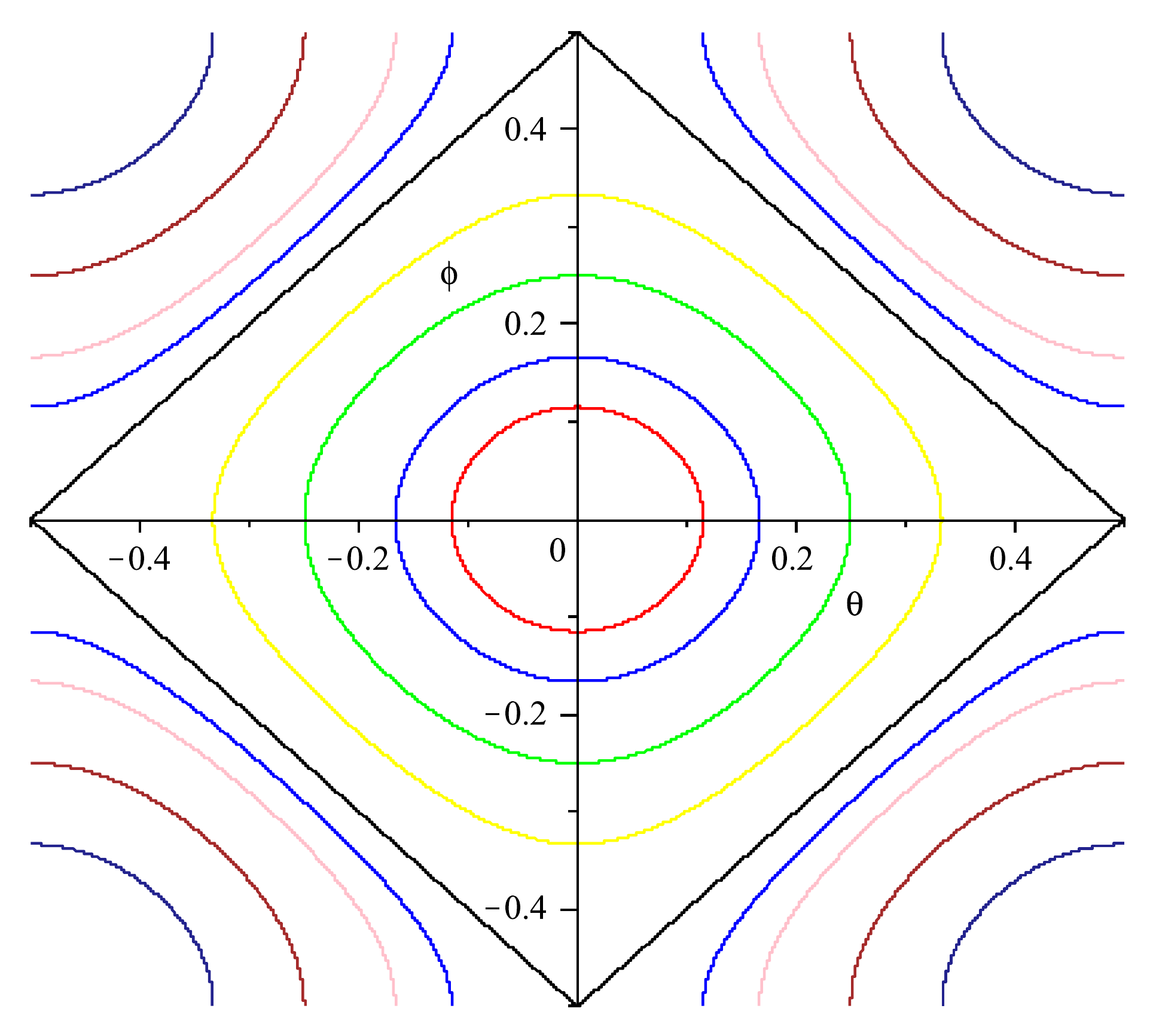}
\caption{\footnotesize{Level sets of the multiplication operator
$\FT \Omega_2\FT^{-1}:f(\theta,\phi)\mapsto (4-2\cos(2\pi \theta)-2\cos(2\pi \phi))f(\theta,\phi)$.}}
 \label{fig:fig2}
\end{figure}



The following theorem says that the component of $H_2$ in the system $\sysbig$ that is reconstructible from the dynamics projected onto $H_1$ is the space of motions that are symmetric with respect to the variable $m$, in other words, motions that are anti-symmetric with respect to the line of coupling of the waveguide cannot excite the waveguide.

\begin{theorem}[part of $H_2$ determined by $H_1$]  The orbit of $\Gamma^\dagger(H_1)$ in $H_2$ under the action of $\Omega_2$ is
\begin{equation*}
  \mathcal{O}_{\Omega_2}(\Gamma^{\dagger}(H_1))=\{\{u_{mn}\} \in H_2: \; u_{mn}=u_{-mn} \; \forall\, m,n \in \mathbb{Z}\}\,.
\end{equation*}
\end{theorem}

\noindent {\emph{Proof.}}
As we have seen, $\Omega_2$ is represented on $L^2([-\onehalf,\onehalf]^2)$ through the Fourier transform $\FT$ by the multiplication operator $T_g f = gf$, where
\begin{equation*}
  g(\theta,\phi) =4-2\cos(2\pi \theta)-2\cos(2\pi \phi).
\end{equation*}
Furthermore, $\FT$ maps the image
\begin{equation*}
  \Gamma^{\dagger}(H_1)=\{ \{u_{mn}\} \in H_2: u_{mn}=0 \text{ for } m\ne 0\}
\end{equation*}
to the subspace $S$ of $L^2([-\onehalf,\onehalf]^2)$ of functions that depend only on $\phi$,
\begin{equation*}
  S=\FT(\Gamma^{\dagger}(H_1))=\{h\in L^2([-\onehalf,\onehalf]^2) : h(\theta,\phi)=h_1(\phi) \text{ for some } h_1 \in L^2([-\onehalf,\onehalf])\}.
\end{equation*}
Since the space of functions $u_{mn}$ in $H_2$ that are symmetric in $m$ is mapped by $\FT$ onto the space of functions in $L^2([-\onehalf,\onehalf]^2)$ that are symmetric in $\theta$, it suffices to prove that
\begin{multline*}
  \text{closure of span }\{\psi(g(\theta,\phi))h_1(\phi) \in L^2([-\onehalf,\onehalf]^2) : \psi\in C(\R), h_1\in L^2([-\onehalf,\onehalf])\} \\
  = \{f\in L^2([-\onehalf,\onehalf]^2) : f(\theta,\phi) = f(-\theta,\phi) \,\forall\, \theta,\phi\in[-\onehalf,\onehalf] \}.
\end{multline*}
Because $g$ is symmetric in $\theta$, all the functions in the space on the left-hand side of this equality are also symmetric in $\theta$.  Thus, the theorem will be proved by showing that
\begin{multline}\label{hello1}
  \text{closure of span }\{\psi(g(\theta,\phi))h_1(\phi) \in L^2([0,\onehalf]\times[-\onehalf,\onehalf]) : \psi\in C(\R), h_1\in L^2([-\onehalf,\onehalf])\} \\
  = L^2([0,\onehalf]\times[-\onehalf,\onehalf]).
\end{multline}
Define
\begin{eqnarray*}
  && \Sigma = \{ \psi(g(\theta,\phi))h_1(\phi) : \psi\in C(\R), h_1\in C([-\onehalf,\onehalf]) \}
       \subset C([0,\onehalf]\times[-\onehalf,\onehalf]), \\
  && \mathfrak{A} = \text{span}(\Sigma).
\end{eqnarray*}
The set $\mathfrak{A}$ is a complex algebra that is closed under conjugation, and it is therefore the algebra generated by the set $\Sigma$.  By taking $\psi$ and $h_1$ to be constant in the definition of $\Sigma$, we find that $\Sigma$ contains the constant functions.  To see that $\Sigma$ separates points, let $(\theta_1,\phi_1)$ and $(\theta_2,\phi_2)$ be distinct points in $[0,\onehalf]\times[-\onehalf,\onehalf]$.  If $\phi_1\not=\phi_2$, then the function $(\theta,\phi)\mapsto\phi$ in $\Sigma$, obtained by taking $\psi$ to be unity and $h_1(\phi)=\phi$ separates these points.  If $\theta_1\not=\theta_2$, then, from the definition of $g$, we see that the function $g(\theta,\phi)$ in $\Sigma$, obtained by setting $\psi(\gamma)=\gamma$ and $h_1(\phi)=1$ separates the points (see Fig.~\ref{fig:fig2}).  By the Stone-Weierstra\ss\ Theorem, $\mathfrak{A}$ is dense in $C([0,\onehalf]\times[-\onehalf,\onehalf])$ in the uniform norm.

Since $C([0,\onehalf]\times[-\onehalf,\onehalf])$ is a dense subset of $L^2([0,\onehalf]\times[-\onehalf,\onehalf])$ in the $L^2$ norm and the $L^2$ norm is bounded by the uniform norm, we conclude that $\mathfrak{A}$ is dense in $L^2([0,\onehalf]\times[-\onehalf,\onehalf])$.  But, since $\mathfrak{A}$ is a subset of the set on the left-hand side of \eqref{hello1}, we have proved \eqref{hello1} and thus established the theorem.
\QED

\section{The time-harmonic scattering problem}

We shall assume the harmonic time-dependent factor $\exp(-i\omega t)$ from now on.

\subsection{Spatial Fourier Harmonics}
Let us consider pseudo-periodic solutions to the problem $(\ref{eq1}),(\ref{eq2})$ with Bloch wave number $\kappa$ in the $n$-direction, which is the direction of the line of coupling.  This means that
\begin{eqnarray*}
  && z_{n+N} = e^{2\pi i\kappa}z_n, \\
  && u_{m,n+N} = e^{2\pi i\kappa}u_{mn}.
\end{eqnarray*}
Such solutions have finite Fourier representations:
\begin{eqnarray*}\label{solution_1D}
  &&z_n=\sum_{\ell=0}^{N-1} c_{\ell} e^{\frac{2 \pi i (\kappa +\ell)}{N}n}, \\
  && u_{mn}=\sum_{\ell=0}^{N-1}(a^{+}_{\ell} e^{2 \pi i \theta_{\ell} m}+a^{-}_{\ell} e^{-2 \pi i \theta_{\ell} m}) e^{2 \pi i \phi_{\ell} n}, \text{ $m\leq0$, } \\
  && u_{mn}=\sum_{\ell=0}^{N-1}(b^{+}_{\ell} e^{2 \pi i \theta_{\ell} m}+b^{-}_{\ell} e^{-2 \pi i \theta_{\ell} m}) e^{2 \pi i \phi_{\ell} n}, \text{ $m\ge0$, }
\end{eqnarray*}
in which $\phi_\ell=(\kappa+\ell)/N$ and $\theta_\ell=\theta_\ell(\kappa,\omega)$ is the $m$-component of the wavevector determined by the dispersion relation \eqref{dispersionrelation} for the operator $\Omega_2$,
\begin{equation*}
  \omega = 4 - 2\cos(2\pi\theta_\ell) - 2\cos(2\pi\phi_\ell).
\end{equation*}
Those values of $\ell$ for which $\theta_\ell$ is real correspond to propagating Fourier harmonics (diffractive orders), and those values of $\ell$ for which $\theta_\ell$ is imaginary correspond to exponential harmonics (evanescent orders).  In the former case, we take $\theta_\ell>0$, and in the latter, we take $i\theta_\ell<0$.  These cases are separated by the case of a linear harmonic $\theta_\ell=0$.

\begin{figure}[ht] \label{N=2,3}
	\centering 
\includegraphics[width=14cm,height=6.5cm]{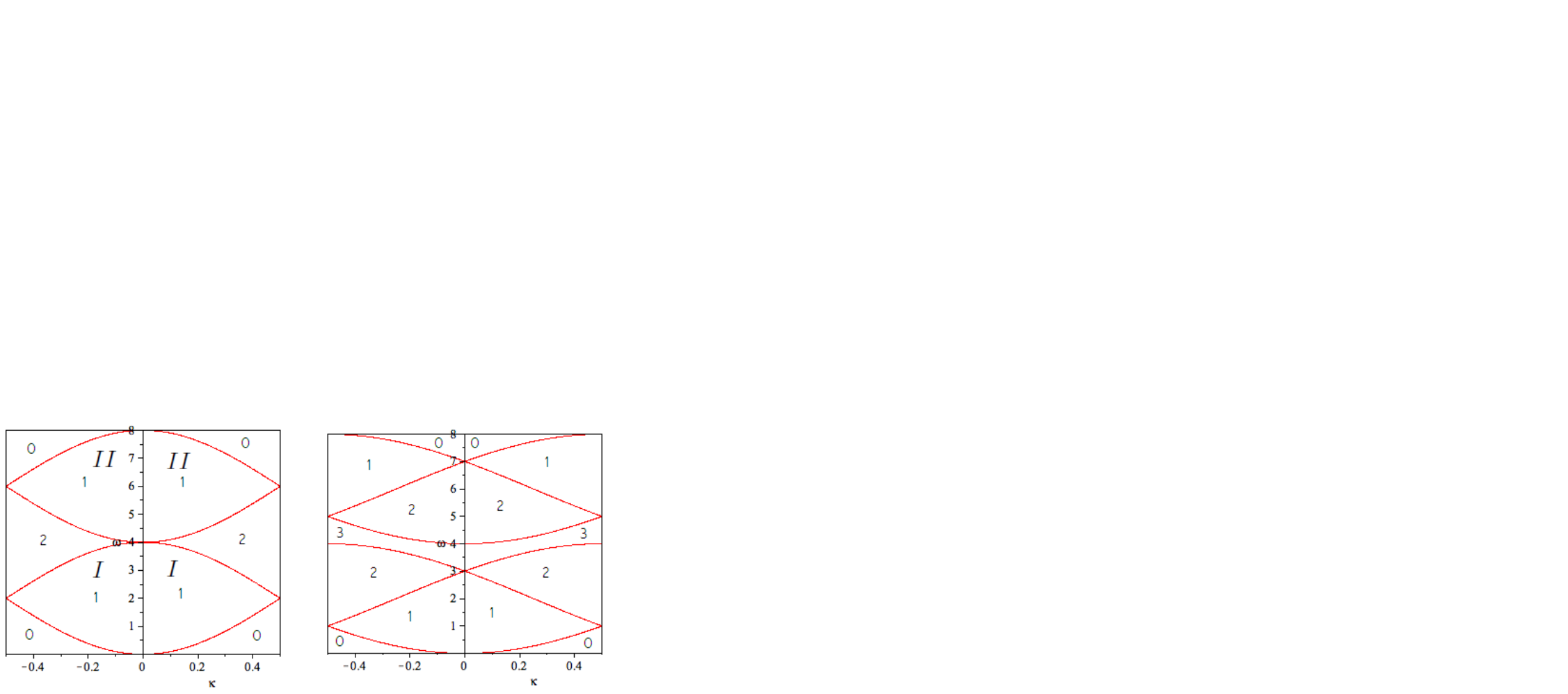}
\caption{\footnotesize{The diagram of $|\mathfrak{P}| $  for $N=2$ (left) and $N=3$ (right). The integers $0$, $1$, $2$, and $3$ represent the number of propagating harmonics.}}
 \label{diagram_2_3}
\end{figure}

\begin{figure}[ht]
	\centering 
\includegraphics[width=14cm,height=6.5cm]{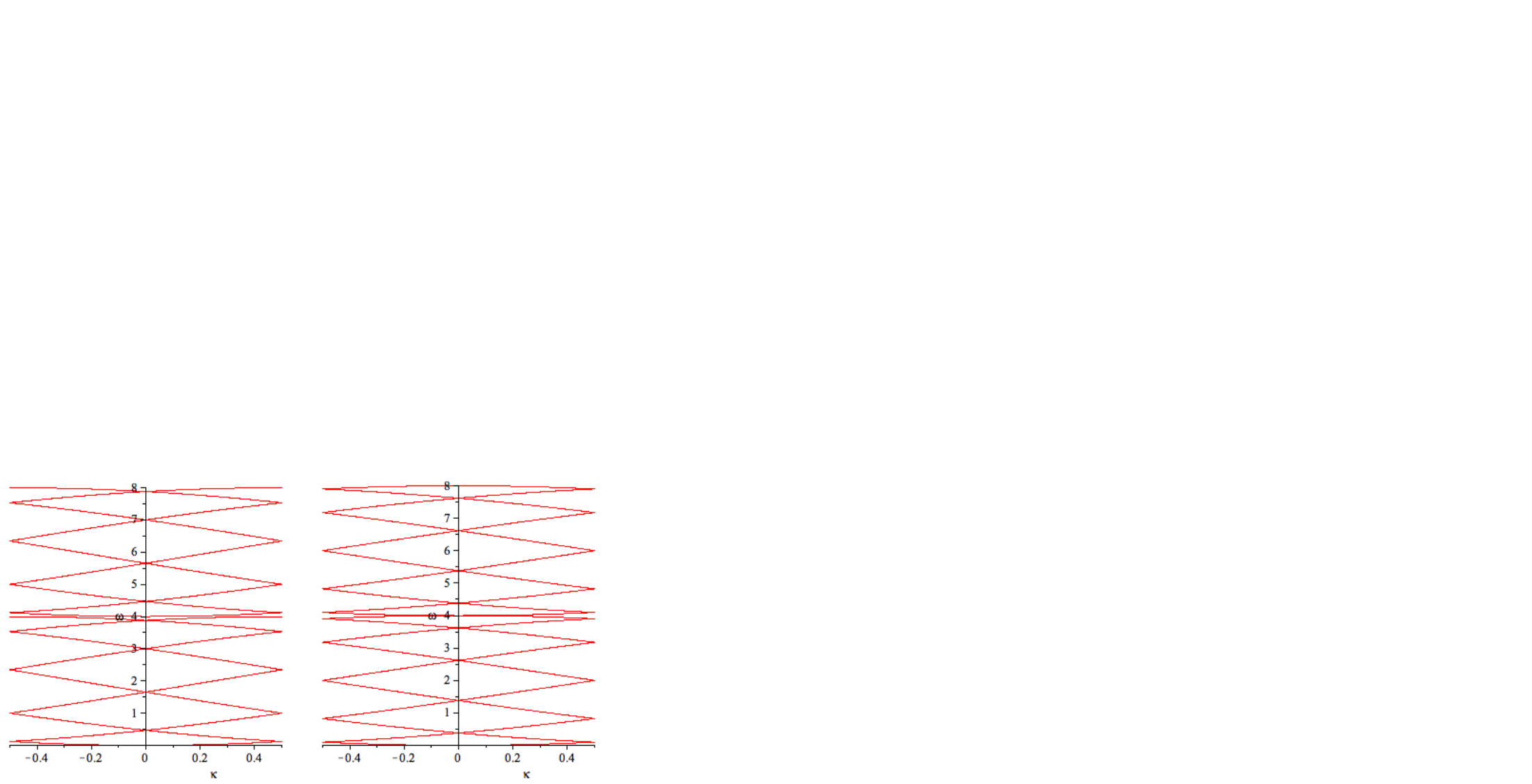}
\caption{\footnotesize{The diagram of $|\mathfrak{P}| $  for $N=9$ (left) and $N=10$ (right).}}
 \label{diagram_9_10}
\end{figure}

Because of the periodicity of the structure, each pseudo-periodic function $u_{mn}$ is characterized by a minimal Bloch wave vector $\kappa$ lying in the first Brillouin zone $\kappa \in [-\onehalf,\onehalf)$. 
The region $[-\onehalf,\onehalf]\times[0,8]$ in $(\kappa,\omega)$-space is divided into sub-regions according to the number of propagating Fourier harmonics. 
For a given pair $(\kappa,\omega)$, let $\mathfrak{P}$ be the set of propagating harmonics,
\begin{equation}
\mathfrak{P} = \mathfrak{P}(\kappa,\omega) =\{\ell:\re(\theta_\ell) \not=0\}.
\end{equation}
Figures~\ref{diagram_2_3} and~\ref{diagram_9_10} show the $(\kappa,\omega)$ regions defined by the order $|\mathfrak{P}|$ of this set as a function of $\kappa$ and~$\omega$.

In the problem of scattering of traveling waves incident upon the waveguide from left and right, we must exclude exponential or linear growth of $\{u_{mn}\}$ in the ambient lattice as $|m| \to \infty$.  Moreover, the energy of the scattered, or diffracted, field must be directed away from the scatter, that is, it must be outgoing.  This notion is made precise by the following definition.

{\definition (outgoing and incoming) \label{out-in} A complex-valued function $\{u_{mn}\}$ is said to be outgoing if there are numbers $\{a_{\ell}\}_{\ell=0}^{N-1}$ and $\{b_{\ell}\}_{\ell=0}^{N-1}$ such that 
\begin{eqnarray}
u_{mn}=\sum\limits_{\ell =0}^{N-1} a_{\ell} e^{-2 \pi i \theta_{\ell}m} e^{2 \pi i \phi_{\ell}n},  \quad m<0 ,\label{out-left}\\
u_{mn}=\sum\limits_{\ell =0}^{N-1} b_{\ell} e^{2 \pi i \theta_{\ell}m} e^{2 \pi i \phi_{\ell}n}, \quad m>0 \label{out-right}.
\end{eqnarray}
The function $\{u_{mn}\}$ is said to be incoming if it admits the expansions
\begin{eqnarray}
u_{mn}=\sum\limits_{\ell =0}^{N-1} a_{\ell} e^{2 \pi i \theta_{\ell}m} e^{2 \pi i \phi_{\ell}n},  \quad m<0, \label{in-left}\\
u_{mn}=\sum\limits_{\ell =0}^{N-1} b_{\ell} e^{-2 \pi i \theta_{\ell}m} e^{2 \pi i \phi_{\ell}n}, \quad m>0 \label{in-right}.
\end{eqnarray}
}
The form of the total field is the sum of an incident field and a scattered one and has the form
\begin{eqnarray}
u_{mn}=\sum\limits_{\ell \in \mathfrak{P}} a_{\ell}^{\mathrm{inc}} e^{2 \pi i \theta_{\ell}m} e^{2 \pi i \phi_{\ell}n}+\sum\limits_{\ell =0}^{N-1} a_{\ell} e^{-2 \pi i \theta_{\ell}m} e^{2 \pi i \phi_{\ell}n},  \quad m\leq0, \label{out-left-total}\\
u_{mn}=\sum\limits_{\ell \in \mathfrak{P}} b_{\ell}^{\mathrm{inc}} e^{-2 \pi i \theta_{\ell}m} e^{2 \pi i \phi_{\ell}n}+\sum\limits_{\ell =0}^{N-1} b_{\ell} e^{2 \pi i \theta_{\ell}m} e^{2 \pi i \phi_{\ell}n} ,\quad m\geq0, \label{out-right-total}
\end{eqnarray}
in which the left-travelling second term for $m\leq0$ consists of the sum of the scattered field to the left of the waveguide and the incident field from the right with coefficients $b_{\ell}^{\mathrm{inc}}$; the field for $m\geq0$ is understood analogously.



{\problem (Scattering problem, $P^{sc}$) Given the coefficients $\{a_{\ell}^{\mathrm{inc}}\}$ and $\{b_{\ell}^{\mathrm{inc}}\}$ of an incident field, 
find a pair of  functions $(z=\{z_n\},u=\{u_{mn}\})$ that satisfies the following conditions:
\begin{eqnarray}
&&\omega z_n=(\Omega_1 z)_n + (\Gamma u)_{n},  \label{1D} \\
&&\omega u_{mn}=(\Gamma^{\dagger}z)_{mn}+(\Omega_2 u)_{mn},  \label{2D}\\
&&(z,u) \text{ are $\kappa$-pseudoperiodic in $n$}, \label{periodic} \\
&&u=u^{\mathrm{inc}}+u^{\mathrm{sc}}, \text{ with $u^{\mathrm{sc}}$ outgoing}, \label{total}
\end{eqnarray}
}
in which
\begin{equation}\label{incident}
  u_{mn}^\mathrm{inc} =
  \sum_{\ell\in\mathfrak{P}} \left[
  a_{\ell}^{\mathrm{inc}} e^{2 \pi i \theta_{\ell}m} +
  b_{\ell}^{\mathrm{inc}} e^{-2 \pi i \theta_{\ell}m}
  \right]\!e^{2 \pi i \phi_{\ell}n} .
\end{equation}
%

\subsection{Conservation of Energy}

Since we seek $\kappa$-pseudo-periodic fields, the scattering problem can restricted to a strip $\mathcal{R}$ containing one period with boundary in the variable $n$,
\begin{equation}
\mathcal{R}=\{(m,n)\in\Z^2: \, -\infty < m < \infty, 0 \le n \le N\}.
\end{equation}
The coupled system admits a law of conservation of energy: The total time-harmonic flux \eqref{flux} out of a truncated region $[m_1,m_2]\times[1,N]$ of the strip $\mathcal{R}$ vanishes; this is stated in the following theorem.

{\theorem \label{equal_energy} Let the frequency $\omega$ and wavenumber $\kappa$ be real.  If the pair $(z_n,u_{mn})$ satisfies the coupled system (\ref{eq1},\ref{eq2}) and has the form
\begin{eqnarray}
&& z_n=\sum\limits_{\ell=0}^{N-1}c_\ell e^{2\pi i \phi_\ell n}, \quad m=0,\label{generalsolution1} \\
&& u_{mn}=\sum_{\ell=0}^{N-1}(a_{\ell}^{-} e^{-2 \pi i \theta_{\ell} m} +a_{\ell}^{+} e^{2 \pi i \theta_{\ell} m}) e^{2 \pi i \phi_{\ell} n}, \quad m\leq0,\label{generalsolution2} \\
&&u_{mn}=\sum_{\ell=0}^{N-1}(b_{\ell}^{-} e^{-2 \pi i \theta_{\ell} m} +b_{\ell}^{+} e^{2 \pi i \theta_{\ell} m}) e^{2 \pi i \phi_{\ell} n}, \quad m\geq0,\label{generalsolution3}
\end{eqnarray}
then
\begin{eqnarray}
\label{flux}
&&\im\left(\sum\limits_{n=1}^{N}(\bar{u}u_{x})_{m_1 n}\right)-\im\left(\sum\limits_{n=1}^{N}(\bar{u}u_{x})_{m_2 n}\right)= \\
&&\label{conservation}
\sum_{\ell \in \mathfrak{P}}\left[(|b_{\ell}^{+}|^2 +|a_{\ell}^{-}|^2)-(|a_{\ell}^{+}|^2 +|b_{\ell}^{-}|^{2})\right]\!\sin{(2 \pi \theta_\ell)}=0,
\end{eqnarray}
where $u_x$ is the forward difference of $u$ in the variable $m$ (see the Appendix).
The fluxes on the upper and lower boundaries cancel identically by pseudo-periodicity.
}

\medskip
\noindent {\emph{Proof:}} We need the following summation-by-parts formula (see the Appendix):
\begin{equation}
\label{Green_1}
\begin{array}{lcl}
\sum\limits_{n=1}^{N}(\bar{z}\Omega_1 z)_{n} &=& \displaystyle
-\frac{k_N\bar{z}_N}{\sqrt{M_N}}(\frac{z_{N+1}}{\sqrt{M_{N+1}}}-\frac{z_N}{\sqrt{M_N}})+\frac{k_0\bar{z}_0}{\sqrt{M_0}}(\frac{z_1}{\sqrt{M_1}}-\frac{z_0}{\sqrt{M_0}})\\
 &  &+ \displaystyle \sum\limits_{n=1}^{N} k_n\left(\frac{\bar{z}_n}{\sqrt{M_n}}-\frac{\bar z_{n-1}}{\sqrt{M_{n-1}}}\right) \left(\frac{z_n}{\sqrt{M_n}}-\frac{z_{n-1}}{\sqrt{M_{n-1}}}\right).
\end{array}
\end{equation}
Upon multiplying $(\ref{eq1})$ by $\bar{z}_n$ and summing over one period, the condition of pseudo-periodicity for $z_n$ and the above formula yield
\begin{equation}
\label{modul1}
0=\sum_{n=1}^{N}\left(\omega|z_n|^{2}-k_n\left|\frac{z_n}{\sqrt{M_n}}-\frac{z_{n-1}}{\sqrt{M_{n-1}}}\right|^2\right)-\sum_{n=1}^{N}\bar{z}_n \gamma_n u_{0n}.
\end{equation}
Similarly, multiplying $(\ref{eq2})$ by $\bar{u}_{mn}$ and using $(\ref{summation})$ we obtain
\begin{equation}
\label{modul2}
0\!=\!\sum_{n=1}^{N}\! \sum\limits_{m=m_1}^{m_2}\!(\omega |u_{mn}|^2\!-\!|\nabla_{\!-}\! u_{mn}|^2)\!+\!\sum\limits_{n=1}^{N}\!(\bar{u} u_x)_{m_2 n}\! -\!\sum\limits_{n=1}^{N}\!(\bar{u} u_x)_{m_1 n}\! -\!\sum\limits_{n=1}^{N}\bar{u}_{0n} \bar{\gamma}_{n} z_n.
\end{equation}
The boundary values at $n=0$ and $n=N$ cancel because of the pseudo-periodicity of $u_{mn}$.
Adding $(\ref{modul1})$ with $(\ref{modul2})$ and taking the imaginary part of the sum leads to the condition $(\ref{flux})$.  Calculation of \eqref{conservation} is straightforward. \QED

\subsection{Formulation in Terms of Fourier Coefficients }

The scattering problem can be reduced to a system of equations for the Fourier coefficients:
\begin{equation}
\label{matrix}
\noindent \begin{cases}
\!\sum\limits_{\ell=0}^{N-1} (a^{-}_{\ell}-b_{\ell}^{+})e^{\frac{2 \pi i \ell}{N}n}=\sum\limits_{\ell=0}^{N-1} (b_{\ell}^{-}-a_{\ell}^{+})e^{\frac{2 \pi i \ell}{N}n},\\
\!\sum\limits_{\ell=0}^{N-1}\! \left(a_{\ell}^{-}\! e^{2 \pi i \theta_\ell}\!-\!b_{\ell}^{+}\!e^{-2 \pi i \theta_\ell}\!-\!\bar{\gamma}_{n} c_\ell\right)e^{\frac{2 \pi i \ell}{N} n}\!=\! \sum\limits_{\ell=0}^{N-1}\!\left(b_{\ell}^{-}\! e^{2 \pi i \theta_\ell}\!-\!a_{\ell}^{+}\! e^{-2 \pi i \theta_\ell}\right) e^{\frac{2 \pi i \ell}{N}n},\\
\!\sum\limits_{\ell=0}^{N-1}\!\left(c_\ell \Big(\omega\! -\!\frac{(k_n+k_{n-1})}{M_n}\!+\!\frac{k_n e^{2 \pi i \frac{\kappa+\ell}{N}}}{\sqrt{M_n M_{n+1}}}\!+\!\frac{k_{n-1} e^{-2 \pi i \frac{\kappa+\ell}{N}}}{\sqrt{M_n M_{n-1}}}\Big)\!-\!\gamma_n b^{+}_{\ell}\right) e^{\frac{2 \pi i \ell}{N}n}\!= \!\gamma_n\! \sum\limits_{\ell=0}^{N-1}\! b_{\ell}^{-}e^{\frac{2 \pi i \ell}{N}n}.
\end{cases}
\end{equation}
or in the matrix form: 
\begin{equation}
\label{matrix1}
\mathbb{B} \overrightarrow{X}=\overrightarrow{F},
\end{equation}
where $\mathbb{B}$ is a $3N\times3N$ matrix, the vector $\overrightarrow{F}$ contains the coefficients $\{a_\ell^+\}$ and $\{b_\ell^-\}$ of the source field, and the vector $\overrightarrow{X}$ represents the coefficients $\{a_\ell^-\}$,  $\{b_\ell^+\}$, of the outgoing field and the coefficients $\{c_\ell\}$ of the field in the waveguide. 

\subsection{Solution of the Scattering Problem}

To prove that the scattering problem always has a solution, it is convenient to work with its variational form.  For this purpose we introduce artificial boundaries at $m=-\mathcal{M}$ and $m=\mathcal{M}$.  At these boundaries, the outgoing condition is enforced through an associated Dirichlet-to-Neumann operator $\mathcal{T}$, which acts on traces on the boundaries $m=\mp \mathcal{M}$ of functions in the pseudo-periodic space ${\mathcal{H}}_{\kappa}(\mathcal{R})$,
\begin{equation}
{\mathcal{H}}_{\kappa}(\mathcal{R})=\{ (z,u) \in \mathcal{H} ( \mathcal{R}): z_N=e^{2\pi i \kappa} z_0, u_{mN}=e^{2\pi i \kappa} u_{m0} \},
\end{equation}
and is defined through the finite Fourier transform as follows.  For any function $v=\{v_{n}\}_{n=0}^{N-1}$, let $\hat v_{\ell}^{\kappa}$ be the $\ell^{\text{th}}$ Fourier coefficient of the function $\{v_{n}e^{-2\pi i \kappa n/N}\}_{n=0}^{N-1}$, that is,
\begin{equation*}
  v_n = \sum_{\ell=0}^{N-1}\hat v_{\ell}^{\kappa} e^{2\pi i(\ell+\kappa)n/N}.
\end{equation*}
Then the map $\mathcal{T}$ is defined through
\begin{equation}
(\widehat{\mathcal{T}v})_{\ell}^{\kappa}=(1-e^{2 \pi i\theta_{\ell}})\hat v_{\ell}^{\kappa}.\end{equation}
The operator $\mathcal{T}$ characterizes the normal forward differences of an outgoing function on the boundary $m=\mp\mathcal{M}$ of the truncated domain in terms of its values there,

\begin{equation}\label{outgoing}
(\partial_{\nu}u +\mathcal{T} u)_{\pm \mathcal{M}n} =0 \; \text{ for $u$ outgoing},
\end{equation}
where 
\begin{eqnarray}
&&(\partial_{\nu}u)_{-\mathcal{M}n}=
-u_{-\mathcal{M}n}+u_{(-\mathcal{M}-1)n}, \\
&&(\partial_{\nu}u)_{\mathcal{M}n}
=u_{(\mathcal{M}+1)n}-u_{\mathcal{M}n}.
\end{eqnarray}
Then using the decomposition $u=u^{\mathrm{sc}}+u^{\mathrm{inc}}$ of the solution to the scattering problem $P^{\mathrm{sc}}$ we obtain
%
\begin{equation}
\begin{array}{rcl}
\partial_{\nu}u+\mathcal{T}u&=&\partial_{\nu}u^{\mathrm{inc}}+\mathcal{T}u^{\mathrm{inc}} \\
 & = &
\begin{cases}
2\sum\limits_{\ell \in \mathfrak{P}}(1-\cos(2\pi \theta_{\ell}))a^{+}_{\ell}e^{2\pi i \theta_{\ell}(- \mathcal{M})}e^{2 \pi i \phi_{\ell} n},  \, m=-\mathcal{M},\\
2\sum\limits_{\ell \in \mathfrak{P}}(1-\cos(2\pi \theta_{\ell}))b^{-}_{\ell}e^{-2\pi i \theta_{\ell} \mathcal{M}}e^{2 \pi i \phi_{\ell} n}, \, m=\mathcal{M}.
\end{cases}
\end{array}
\end{equation}
Thus we are led to the following problem set in the bounded domain $\mathcal{R}^{\mathcal{M}}$ of $\Z^2$: 
\begin{equation}
\mathcal{R}^{\mathcal{M}}=[-\mathcal{M},\mathcal{M}]\times[0,N].
\end{equation}

{\problem (Scattering problem reduced to a bounded domain, $P^{\mathrm{sc}}_{\!\mathcal{M}}$) Find $(z,u)$ in $\mathcal{H}(\mathcal{R}^{\mathcal{M}})$ such that 
\begin{eqnarray}
&&\omega z_n=(\Omega_1 z)_n + (\Gamma u)_{n} \;\text{ for } 0\leq n\leq N,  \label{1DM} \\
&&\omega u_{mn}=(\Gamma^{\dagger}z)_{mn}+(\Omega_2 u)_{mn} \;\text{ for } (m,n)\in\mathcal{R}^\mathcal{M},  \label{2DM}\\
&&(z,u) \text{ are $\kappa$-pseudoperiodic in $n$}, \label{periodicM} \\
&&\partial_{\nu}u+\mathcal{T}u=\partial_{\nu}u^{\mathrm{inc}}+\mathcal{T}u^{\mathrm{inc}} \text{ on $m=\mp \mathcal{M}$ } .\label{totalM}
\end{eqnarray}}
\smallskip
Problems $P^{\mathrm{sc}}$ and  $P^{\mathrm{sc}}_{\!\mathcal{M}}$ are equivalent in the sense of the following theorem. 

{\theorem If $(z,u)$ is a solution of $P^{\mathrm{sc}}$ such that $(\tilde{z},\tilde{u})=(z,u)\vert_{R^{\mathcal{M}}} \in \mathcal{H}(R^{\mathcal{M}})$, then $(\tilde{z},\tilde{u})$ is a solution of $P^{\mathrm{sc}}_{\!\mathcal{M}}$. Conversely, if $(\tilde{z},\tilde{u})$ is a solution of $P^{\mathrm{sc}}_{\!\mathcal{M}}$, it can be extended uniquely to a solution $(z,u)$ of $P^{\mathrm{sc}}$.
}

\smallskip
\noindent {\emph{Proof:}} The first part of the theorem holds because the condition $(\ref{totalM})$ is equivalent to $(\ref{total})$. Conversely, if $(\tilde{z},\tilde{u})$ is a solution of $P^{\mathrm{sc}}_{\!\mathcal{M}}$, then by \ref{totalM}, the difference $\tilde u^\mathrm{sc} := \tilde{u}-u^\mathrm{inc}$ satisfies the Dirichlet-Neumann relation \eqref{outgoing} that characterizes outgoing fields and can therefore be extended to an outgoing field $u^\mathrm{sc}$.   The field $u=u^\mathrm{inc} + u^\mathrm{sc}$, by the definition of its parts, satisfies (\ref{1D},\ref{2D}) outside of $\mathcal{R}^\mathcal{M}$.
\QED

 
\smallskip
A variational form of the scattering problem, which is analogous to the weak formulation for partial differential equations, is obtained from the summation by parts formulas in the Appendix.
With the notation $\partial_y$ and $\partial_{\bar y}$ for the forward and backward differences in the $n$-variable, one can write $\Omega_1 = -M^\half\partial_{\bar y}K\partial_y M^\half$, where $(Kv)_n = k_nv_n$, and use the substitution $v\mapsto K\partial_y M^\half z$ and $w\mapsto M^\half\bar w$ in \eqref{discrete1d} to obtain the first equation below.  The second is obtained by using equation \eqref{summation}.  We use the notation $\nabla_{\!-} = (\partial_{\bar x},\partial_{\bar y})$.

It is straightforward to prove that problems $P^{\mathrm{sc}}_{\!\mathcal{M}}$ and $P^{\mathrm{sc}}_{\mathrm{\!\!var}}$ are equivalent.

{\problem(Scattering Problem, variational form, $P^{\mathrm{sc}}_{\mathrm{\!\!var}}$) Find a function $(z,u)\in {\mathcal{H}}_{\kappa}(R^{\mathcal{M}})$ such that
\begin{equation}
\label{wfcsys}
\begin{array}{ll}
\!\sum\limits_{n=0}^{N-1}(K\partial_y M^{-\half}z)_n (\partial_y M^{-\half}\bar{w})_n +\sum\limits_{n=1}^{N}((\Gamma u)_n \bar{w}_{n}-(\omega z)_{n}\bar{w}_n)=0,\\
\!\!\sum\limits_{n=1}^{N}\!\sum\limits_{m=-\mathcal{M}}^{\mathcal{M}}\!(\omega u\bar{v}\!\!-\!\!\nabla_{\!-}\bar{v}\,\nabla_{\!-}u\!\!-\!\!(\Gamma^{\dagger}z)\bar{v})_{mn})\!\!-\!\!
\sum\limits_{n=1}^{N}\!(\bar{v}_{(-\mathcal{M}-1)n}(\mathcal{T}u)_{\!-\mathcal{M}n}\!\!-\!\!(\bar{v}\mathcal{T}u)_{\mathcal{M}n})\\
\hfill\!=\!
-\sum\limits_{n=1}^{N}(\bar{v}_{(-\mathcal{M}-1)n}(\partial_{\nu}u^{\mathrm{inc}}\!+\!\mathcal{T}u^{\mathrm{inc}}))_{-\mathcal{M}n}\!-\!\sum\limits_{n=1}^{N}(\bar{v}(\partial_{\nu}u^{\mathrm{inc}}\!+\!\mathcal{T}u^{\mathrm{inc}}))_{\mathcal{M}n}.
\end{array}
\end{equation} 
for all $w\in \mathcal{H}_{\kappa}(\mathcal{R}^{\mathcal{M}})$ and $\kappa$-pseudoperiodic $v=\{v_n\}$.
}

{\theorem (Equivalence of $P^{\mathrm{sc}}_{\!\mathcal{M}}$ and $P^{\mathrm{sc}}_{\mathrm{\!\!var}}$) If $(z,u)\in\mathcal{H}_{\kappa}(\mathcal{R}^{\mathcal{M}})$ satisfies the scattering problem $P^{\mathrm{sc}}_{\!\mathcal{M}}$, then $(z,u)$ satisfies $P^{\mathrm{sc}}_{\mathrm{\!\!var}}$. Conversely, if $(z,u)$ satisfies $P^{\mathrm{sc}}_{\mathrm{\!\!var}}$ for any $(w,v)\in\mathcal{H}_{\kappa}(\mathcal{R}^{\mathcal{M}})$, then $(z,u)$ satisfies $P^{\mathrm{sc}}_{\!\mathcal{M}}$ also.
}

\smallskip
The scattering problem always has a solution, even if it is not unique.  Non-uniqueness occurs when the structure supports a guided mode, as we discuss in the next section.  The reason for the existence of a solution of the scattering problem in the presence of guided modes lies in the orthogonality of guided modes to incident plane waves: the former possess only evanescent harmonics, while the latter possess only propagating harmonics.  This fact will be important in the analysis of resonant amplitude enhancement, and it has its analog in continuous problems of scattering of waves by open periodic waveguides \cite[Thm. 3.1]{Bonnet-BeStarling1994}

{\theorem  \label{solution} The problem $(\ref{wfcsys})$ always has a solution.} 

\smallskip
\noindent {\emph{Proof:}} Let us rewrite $(\ref{wfcsys})$ in the concise inner product form 
\begin{equation}
\label{simple}
\langle AY,V \rangle=\langle F,V \rangle \quad \text{for all } V,
\end{equation}
where $Y=(z,u)$, $F=(0,f)$ and $V=(w,v)$.
We use the Fredholm alternative, namely, that \eqref{simple} has a solution $(z,u)$ if and only if $\langle F,V \rangle=0$ for all $V \in \text{Null}(A^{*})$, or, in other words,
\begin{equation}
\label{other}
\langle F,V \rangle=0 \text{ for all } V \text{ such that } \langle AY,V \rangle=0 \text{ for all } Y .
\end{equation}   
Any function $V$ satisfying $\langle AY,V \rangle=0$ for all $Y$ satisfies $\langle AV,V \rangle=0$ as well.  By Theorem~\ref{equal_energy} it follows that $v$ contains only evanescent harmonics (or linear ones for threshold values of $(\kappa,\omega)$) in its Fourier series, that is
\begin{equation}
\label{rep}
v_{mn}=
\begin{cases}
\sum\limits_{\ell \in \bar{\mathfrak{P}}}v_{\ell}^{-}e^{-2\pi i \theta_{\ell}m}e^{2\pi i \phi_{\ell}n}  & \text{ for $m\le0$ },\\
\sum\limits_{\ell \in \bar{\mathfrak{P}}}v_{\ell}^{+}e^{2\pi i \theta_{\ell}m}e^{2\pi i \phi_{\ell}n}  & \text{ for $m\ge0$ },
\end{cases}
\end{equation}
where $\bar{\mathfrak{P}} = {1,\dots,N}\setminus\mathfrak{P}$.  Using the orthogonality of the Fourier harmonics, we obtain
\begin{equation}
\label{check}
\begin{array}{rcl}
\langle F,V \rangle&\!\!=\!\!&\langle f,v \rangle\\
 &\!\!=\!\!&\!\! -2\!\!\sum\limits_{n=1}^{N}(\!(\!\sum\limits_{\ell \in \bar{\mathfrak{P}}}\!\bar{v}_{\ell}^{-}e^{2\pi i \theta_{\ell}(-\mathcal{M}-1)}e^{-2\pi i \phi_{\ell}n})
(\!\sum\limits_{\ell' \in \mathfrak{P}}\!(\!1-\cos(2 \pi \theta_{\ell'})\!)a_{\ell'}^{+}e^{2\pi i \theta_{\ell'}(-\mathcal{M})}e^{2\pi i \phi_{\ell'}n})\!)\\
 & \!\!&\!\! -2\!\!\sum\limits_{n=1}^{N}(\!(\!\sum\limits_{\ell \in \bar{\mathfrak{P}}}\bar{v}_{\ell}^{+}e^{-2\pi i \theta_{\ell}\mathcal{M}}e^{-2\pi i \phi_{\ell}n})
  (\sum\limits_{\ell' \in \mathfrak{P}}\!(\!1-\cos(2 \pi \theta_{\ell'})\!)b_{\ell'}^{-}e^{-2\pi i \theta_{\ell'}\mathcal{M}}e^{2\pi i \phi_{\ell'}n})\!)\\
 &\!\!=\!\!&\!-\!2\!\!\sum\limits_{\ell \in \bar{\mathfrak{P}}\atop \ell'\in \mathfrak{P}}\!(\!(\bar{v}_{\ell}^{-}\!e^{2\pi i \theta_{\ell}(-\!\mathcal{M}\!-\!1)}(\!1\!\!-\!\cos(\!2\pi \theta_{\ell'}\!)\!)a_{\ell'}^{+}e^{2\pi i \theta_{\ell'}(\!-\!\mathcal{M})}\!)
(\!\sum\limits_{n=1}^{N}\!\!e^{2 \pi i(\phi_{\ell'}-\phi_{\ell})n})\!)\\
 &\!\! \!\!&\!-\!2\!\!\sum\limits_{\ell \in \bar{\mathfrak{P}}\atop \ell'\in \mathfrak{P}}\!
(\!(\bar{v}_{\ell}^{+}\!e^{-2\pi i \theta_{\ell}\mathcal{M}}(\!1\!\!-\!\cos(\!2\pi \theta_{\ell'}\!)\!)b_{\ell'}^{-}e^{-2\pi i \theta_{\ell'}\mathcal{M}})
(\!\sum\limits_{n=1}^{N}\!\!e^{2 \pi i(\phi_{\ell'}-\phi_{\ell})n})\!)=0.
\end{array}
\end{equation}
Therefore there exists a solution $(z,u)$ to the problem $(\ref{wfcsys})$.
\QED

\section{Guided modes}

A \emph{guided mode} is a nontrivial solution of problem $P^\mathrm{sc}$ in which the incident field is set to zero.

{\problem (Guided mode problem, $P^{gm}$) Find a pair of  functions $(z=\{z_n\},u=\{u_{mn}\})$ that satisfies the following conditions:
\begin{eqnarray}
&&\omega z_n=(\Omega_1 z)_n + (\Gamma u)_{n},  \label{1Dgm} \\
&&\omega u_{mn}=(\Gamma^{\dagger}z)_{mn}+(\Omega_2 u)_{mn},  \label{2Dgm}\\
&&(z,u) \text{ are $\kappa$-pseudoperiodic in $n$}, \label{periodicgm} \\
&&u \; \text{is outgoing}. \label{totalgm}
\end{eqnarray}
}
Because of conservation of energy relation \eqref{conservation}, a generalized guided mode supported by the structure at a real pair $\kw$ possesses no propagating harmonics and is therefore exponentially decaying as $|m|\to\infty$.  It may be called a true guided mode.  

In a real $\kw$ region in which all harmonics are evanescent (see the region labelled ``0" in Fig.~\ref{diagram_2_3}), there are, for appropriate values of $M_i$ and $\gamma_i$, dispersion curves defining the locus of $\kw$-pairs that support a guided mode.  These are robust in the sense that, as $\kappa$ is perturbed, the guided mode persists albeit at a different frequency.  Physically speaking, energy cannot radiate away from the waveguide because there are no propagating harmonics available for transporting the energy (the energy flux of all evanescent harmonics is along the waveguide).

The situation is different in a $\kw$-region of, say, one propagating harmonic (the region labeled ``1" in Fig.~\ref{diagram_2_3}), as it typically contains no dispersion curves for guided modes.  Nevertheless, a structure may support a guided mode at an \emph{isolated} $\kw$-pair in this region.  Such a guided mode is nonrobust with respect to perturbations of $\kappa$.  The physical idea is that the propagating harmonics, which appear with nonzero coefficients in a typical solution (\ref{generalsolution1}--\ref{generalsolution3}) are carriers of incoming and outgoing radiation.  But special conditions that allow these coefficients to vanish at some frequency, thereby creating a guided mode, may be arranged by tuning the structure (masses and coupling constants) and the wavenumber to specific parameters.  A perturbation of $\kappa$ from the value $\kappa_0$ that supports the guided mode will destroy these special conditions.  In physical terms, one often describes the destruction of a guided mode as the coupling or interaction of the mode with radiative fields.  It is this interaction that causes resonant scattering behavior and transmission anomalies.

\subsection{Generalized guided modes}

A rigorous analysis of resonance near nonrobust guided modes requires an extension of the scattering problem ($P^\mathrm{sc}$) and the problem of guided modes ($P^\mathrm{gm}$) to complex $\kappa$ and $\omega$ in a vicinity of the guided mode pair $\kwz$.  There arises a complex dispersion relation $D\kw=0$ in $\C^2$ describing the locus of generalized guided modes, given by the zero set of the determinant of the matrix $\mathbb{B}$ in \eqref{matrix1}.  From this point of view, an isolated guided-mode pair in real $\kw$-space is the intersection in $\C^2$ between the real plane and the dispersion relation.
We call solutions of Problem $P^{gm}$ for complex $\kappa$ or $\omega$ ``generalized guided modes"; they are foundational to the theory of leaky modes, as discussed, for example, in \cite{HausMiller1986,PaddonYoung2000,PengTamirBertoni1975,TikhodeevYablonskiMuljarov2002}.

Let us consider real values of $\kappa$ and examine how the nature of the Fourier harmonics changes when $\omega$ is allowed to assume a small imaginary part.  Suppose that $\theta_\ell>0$ for some real $\omega=\omega_R$, that is, the $\ell^\text{th}$ harmonic is propagating.  Then for $\omega=\omega_R+i\omega_I$, where $\omega_I$ is small, $\theta_\ell$ also attains a small imaginary part, $\theta_\ell=\theta^{R}_{\ell}+i\theta_{\ell}^{I}$.
The dispersion relation gives
$$
2\cos(2\pi (\theta^{R}_{\ell}+i\theta^{I}_{\ell}))=4-\omega_{R}-i\omega_{I}-2\cos(2\pi \phi_\ell).
$$
Taking the imaginary part,
one finds
$$
\sin(2 \pi \theta^{R}_{\ell})(e^{2 \pi \theta^{I}_{\ell}}-e^{-2 \pi \theta^{I}_{\ell}})=\omega_{I}.
$$
Thus, if $\omega_{I}$ is negative (and sufficiently small), $\theta^{I}_{\ell}$ is also negative.  This means that the outgoing Fourier harmonic $e^{2\pi i(\phi_\ell n+\theta_\ell |m|)}e^{-i\omega t}$ decays in time but grows in space as $|m|\to\infty$ whereas the incoming one decays in space and time.  Conversely, if $\omega_{I}$ is a small positive number, then $\theta^{I}_{\ell}>0$ and the $\ell^\text{th}$ outgoing harmonic grows in time and decays in space and the incoming one grows in space and time.
The evanescent (resp. growing) harmonics, for which $\theta_\ell<0$, remain evanescent (resp. growing) under small imaginary perturbations of $\omega$.

The following theorem is the discrete analog of Theorem 5.2 in \cite{ShipmanVenakides2003} for photonic crystal slabs, which states that generalized guided modes occur only for $\im(\omega)\leq0$ and that such a mode is a true evanescent one if and only if $\im(\omega)=0$.

{\theorem \label{nonzero im} Suppose that $(\{z_n\},\{u_{mn}\})$ is a nontrivial solution to the homogeneous (sourceless) problem $P^{gm}$. Then $\im(\omega) \le 0$. In addition, $|u_{mn}| \to 0$ as $|m| \to \infty$ if and only if $\im(\omega)=0$.}

\smallskip
\noindent {\emph{Proof:}} Adding $(\ref{modul1})$ and $(\ref{modul2})$ and taking the imaginary part yields 
\begin{equation}
\label{sixty-fourth}
\im(\omega) \Big(\sum_{n=1}^{N} |z_{n}|^2 + \sum_{n=1}^{N}\sum_{m=m_1}^{m_2} |u_{mn}|^2\Big)=\im\Big(\sum_{n=1}^{N}(\bar{u}u_{x})_{m_1 n} -\sum_{n=1}^{N}(\bar{u}u_{x})_{m_2 n}\Big).
\end{equation}
\normalcolor
If the field decays as $|m|\to\infty$, then the right-hand side of \eqref{sixty-fourth} tends to zero as $-m_1$ and $m_2$ tend to $\infty$, and thus the left-hand side vanishes.  Since the field is nontrivial, we obtain $\im(\omega)=0$.  Conversely, if $\im(\omega)=0$, then Theorem~\ref{equal_energy} applies and we find that, since the coefficients $a_\ell^+$ and $b_\ell^-$ ($\ell\in\mathfrak{P}$) of the incident field vanish, so also must $a_\ell^-$ and $b_\ell^+$ for $\ell\in\mathfrak{P}$.  Thus $u_{mn}$ decays exponentially as $|m|\to\infty$.

If $\im({\omega})>0$, then, as we have mentioned, the generalized outgoing Fourier harmonics of the field decay as $|m|\to\infty$.  Since there are no incoming harmonics by assumption, the right-hand side of \eqref{sixty-fourth} decays as $|m|\to\infty$, so the left-hand side vanishes, and we must have $\im(\omega)=0$.  We conclude that, for all solutions of the homogeneous problem $P^\text{gm}$, it is necessary that $\im(\omega)\leq0$.
\QED
\medskip

\subsection{The spectrum of $\Omega_\kappa$}

As we have mentioned earlier, the operator $\Omega$ is a direct integral of the self-adjoint operators $\Omega_\kappa$.  Since the latter acts on the space of $\kappa$-pseudo-periodic functions $u_{mn}$, its domain may be taken to be $\ell^2(\mathcal{R})$ with the pseudo-periodic boundary condition $u_{mN}=e^{2\pi i\kappa}u_{m0}$.  The spectrum of $\Omega_\kappa$ consists of a continuous part and a set of eigenvalues.  The continuous spectrum is the $\omega$-coordinates of the intersection of the region of at least one propagating harmonic in Figs.~\ref{diagram_2_3} and \ref{diagram_9_10} with the vertical line at fixed $\kappa$.  The eigenvalues are the real values of $\omega$ for which $\mathbb{B}$ has a nullspace, that is, the frequencies that support a true guided mode for the given value of $\kappa$.  These frequencies may be either in the region of no propagating harmonics or embedded in the continuous spectrum.

In the context of spectral theory, an isolated point $\kwz$ of the dispersion relation $D\kw=0$, within a region of at least one propagating harmonic, corresponds to an embedded eigenvalue of the operator $\Omega_\kappa$ that dissolves into the continuous spectrum as $\kappa$ is perturbed from $\kappa_0$.  The associated destruction of the guided mode is associated with transmission resonance.  This phenomenon is akin to the quantum-mechanical resonances of the noble gases, in which an embedded bound state of the idealized atom with no interaction between the electrons is destroyed when this interaction is initiated, \cite{Fano1961}, \cite[\S XXII.6]{ReedSimon1980d}.

\subsection{Existence of Guided Modes}



Let us consider the existence of guided modes in the case of period $N\!=\!2$, which is a minimal model for the phenomenon of anomalous transmission.  Depending on the values of $\kappa$ and $\omega$, there are either two, one, or no propagating harmonics, as depicted in Fig.~\ref{diagram_2_3}.  We are interested in regions $I$ and $II$, in which there is exactly one propagating and one evanescent harmonic.  It is this region in which one may encounter a nonrobust guided mode and associated scattering resonance.  

In the region $I$, the exponent $\theta_0$ corresponds to a propagating harmonic, whereas $\theta_1$ corresponds to an evanescent one.  The coefficients of the propagating harmonic in the Fourier decomposition of the solution to the guided mode problem ($P^\mathrm{gm}$) are forced to be zero. The corresponding unknown vector in equation \eqref{matrix1} is $\vec{X}=(a_0^{-}=0,b_0^{+}=0,c_0,a_1^-,b_1^+,c_1)$.  The condition that the values $c_0$,~$a_1^-$, $b_1^+$, and $c_1$ not vanish simultaneously in the system $(\ref{matrix})$ yields two equations that characterize a guided mode
\begin{multline}
\label{criterion1}
\frac{(\bar{\gamma}_1-\bar{\gamma}_0)}{(\bar{\gamma}_0+\bar{\gamma}_1)}\left((k_0+k_1)(\frac{1}{M_1}-\frac{1}{M_0})+\frac{2i\sin(\pi\kappa)}{\sqrt{M_0M_1}}(k_0-k_1)\right) \\ -\frac{\bar{\gamma}_0\bar{\gamma}_1(\gamma_0+\gamma_1)}{(\bar{\gamma}_0+\bar{\gamma}_1)i \sin(2\pi \theta_1)}+2\omega+(k_0+k_1)(-\frac{1}{M_0}-\frac{1}{M_1}-\frac{2 \cos(\pi \kappa)}{\sqrt{M_0 M_1}})=0,\\
\vspace{-1ex} 
\end{multline}
\begin{multline}
\label{criterion2}
\frac{(\bar{\gamma}_1-\bar{\gamma}_0)}{(\bar{\gamma}_0+\bar{\gamma}_1)}\left(2\omega+(k_0+k_1)(\frac{2 \cos(\pi \kappa)}{\sqrt{M_0 M_1}}-\frac{1}{M_0}-\frac{1}{M_1})\right)\\ +\frac{\bar{\gamma}_0\bar{\gamma}_1(\gamma_1-\gamma_0)}{(\bar{\gamma}_0+\bar{\gamma}_1)i \sin(2\pi \theta_1)}+(k_0+k_1)(\frac{1}{M_1}-\frac{1}{M_0})+\frac{2i\sin(\pi\kappa)(k_1-k_0)}{\sqrt{M_0M_1}}\!=0 ,
\end{multline}
where $\sin(2\pi \theta_1)=\sqrt{1-(2-\frac{\omega}{2}+\cos(\pi \kappa))^2}$.
For any real pair $\kwz$ satisfying (\ref{criterion1},\ref{criterion2}) with $\kappa_0\in[-\onehalf,\onehalf]$, we must also ensure that $\omega_0\in(2-2\cos(\pi\kappa_0),2+2\cos(\pi\kappa_0))$, that is, that $\kwz$ lies within region $I$.

Region $II$ can be treated similarly.  There, the roles of $\theta_0$ and $\theta_1$ are switched and one seeks nonzero solutions for $\overrightarrow{X}=(a_0^{-},b_0^{+},c_0,a_1^-=0,b_1^+=0,c_1)$.
 
\begin{figure}[ht] 
	\centering 
\includegraphics[width=14.5cm,height=5.6cm]{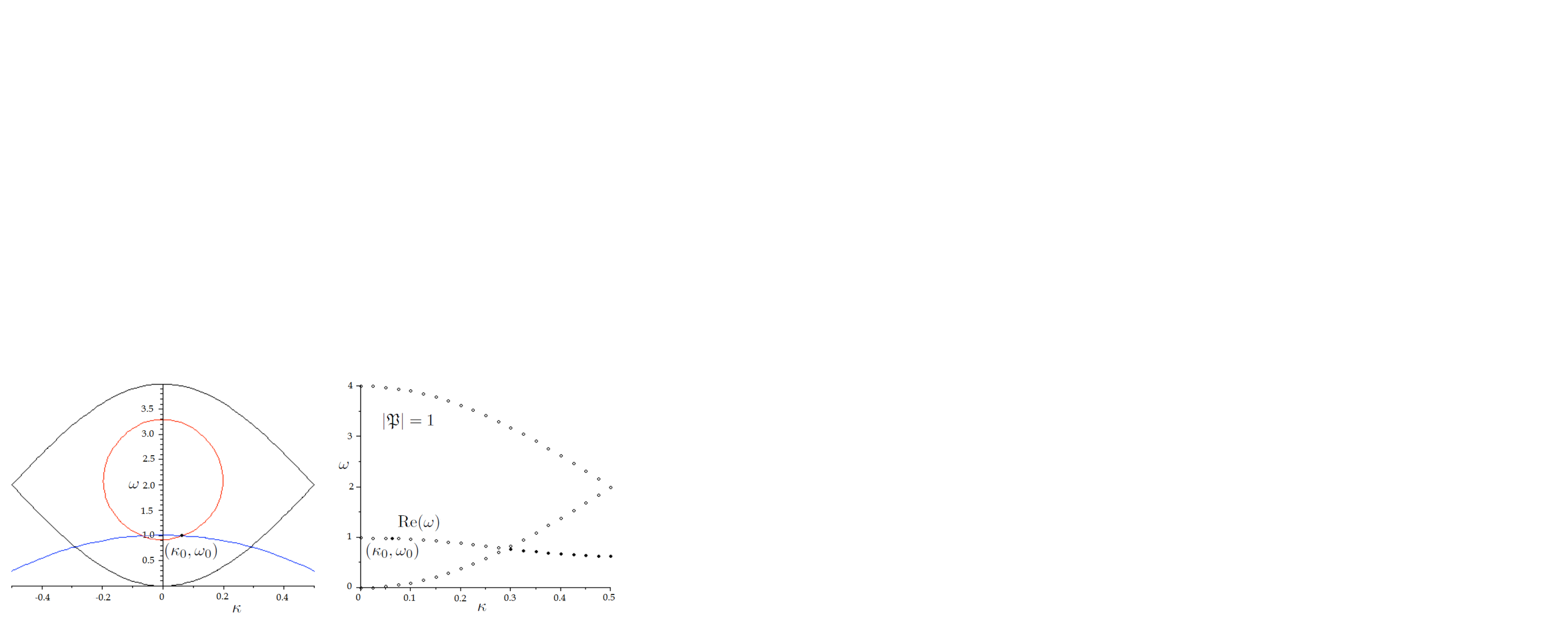}
\caption{\footnotesize{Example for $M_0=2$, $M_1=1$, $k_0=k_1=1$, $\gamma_0=1$, and $\gamma_1=7$. Left: The intersection of the two relations (\ref{criterion1},\ref{criterion2}) (in red and blue) is the locus of guided modes, in this case $(\kappa_0=\pm0.0616,\omega_0=0.9792)$.   A region of one propagating harmonic is outlined in black.  Right: The unfilled circles outline the region of one propagating harmonic and indicate the real part of the dispersion relation within this region.  The pair $\kwz$ of a guided mode in this region is represented as a solid black dot, as is the real dispersion relation for guided modes in the region of no propagating harmonics.}}
 \label{Cross}
\end{figure}

 Figures \ref{Cross} and \ref{Touch} show the locus of the solutions of (\ref{criterion1},\ref{criterion2}) in the $\kw$ plane for different choices of the structural parameters of the system.  The points of intersection come in $\pm\kappa$ pairs when the $\kappa$-value of the corresponding guided mode is nonzero (Fig.~\ref{Cross}).  Fig.~\ref{Touch} shows the case of a guided mode at $\kappa=\kappa_0=0$, which, as the system parameters are perturbed, either bifurcates into two modes at $\pm\kappa_0\not=0$ or disappears altogether.  We will examine this bifurcation in more detail in Section~\ref{sec:resonantscattering}.
  

For certain structures, one can prove the absence of non-robust guided modes.

{\theorem For the coupled system of period two with $M_0\ne M_1$, $k_0=k_1=k\ne0$,  and $\pm\gamma_0=\gamma_1=\gamma\ne0$, there is no guided-mode pair $(\kappa,\omega)$ in the subregion of $[-\onehalf,\onehalf]\times [0,8]$ that admits at least one propagating harmonic.} 

\medskip
\noindent {\emph{Proof:}} Suppose the real pair $(\kappa,\omega)$ admits a solution $\{u_{mn},z_n\}$ of Problem $P^{gm}$. By Theorem~\ref{nonzero im}, $|u_{mn}| \to 0$ as $|m| \to \infty$, and thus the solution contains only evanescent harmonics.  Given that $\kw$ admits at least one propagating harmonic, we must have $|\mathfrak{P}|=1$ (see Fig.~3, left).  In the region $I$ defined by $\kappa \in [-\onehalf,\onehalf]$, $\omega \in [2-2\cos(\pi\kappa),2+2\cos(\pi\kappa)]$, with propagating and evanescent harmonics corresponding to $\theta_0\in\R$ and $\theta_1\in i\R$, respectively, the solution has the form
\begin{equation}
\begin{array}{l}
u_{mn}=a_{1}^{-} e^{-2 \pi i \theta_1 m} e^{2 \pi i \phi_1 n}, \quad m \leq 0, \\
u_{mn}=b_{1}^{+} e^{2 \pi i \theta_1 m} e^{2 \pi i \phi_1 n}, \quad m \geq 0, \\
z_n=c_0 e^{2 \pi i \phi_0 n}+c_1e^{2 \pi i \phi_1 n},
\end{array}
\end{equation}
\noindent with $a_0^{-}=b_0^{+}=0$ corresponding to the propagating harmonics.  Using the restrictions on $k_i$ and $\gamma_i$ given in the Theorem, the system $(\ref{matrix})$ with zero right-hand side can be reduced to derive $c_0=c_1=0$ and $a_1^-=b_1^+=0$. 
Region $II$ is handled similarly.
 \QED

\begin{figure}[ht]  
	\centering 
\includegraphics[width=14.5cm,height=5.9cm]{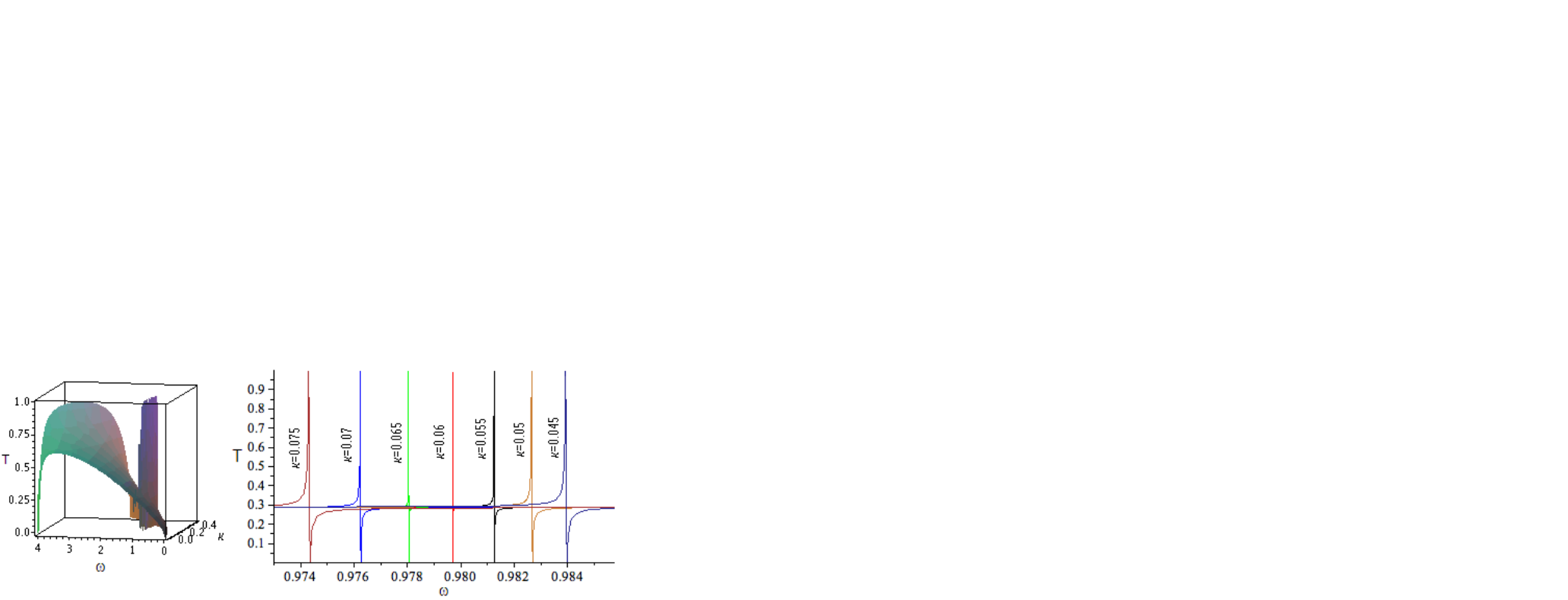}
\caption{\footnotesize{Left: The transmission coefficient for $M_0=2$, $M_1=1$, $k_0=k_1=1$, $\gamma_0=1$, $\gamma_1=7$.  {There is a guided mode at $(\kappa_0,\omega_0)\approx(0.0616, 0.9792)$.} Right: Refined graphs of $T$ {\itshape vs.} $\omega$ for various values of $\kappa$ near $\kappa_0$.}
Observe that $\ell_1\not=0$ in the formula of Theorem~\ref{error_T} for this example, so that the center of the resonance varies linearly as a function of $\kappa-\kappa_0$, whereas the width varies quadratically.}
 \label{Resolution_Cross}
\end{figure}

\begin{figure}[ht] 
	\centering 
\includegraphics[width=14.5cm,height=5.9cm]{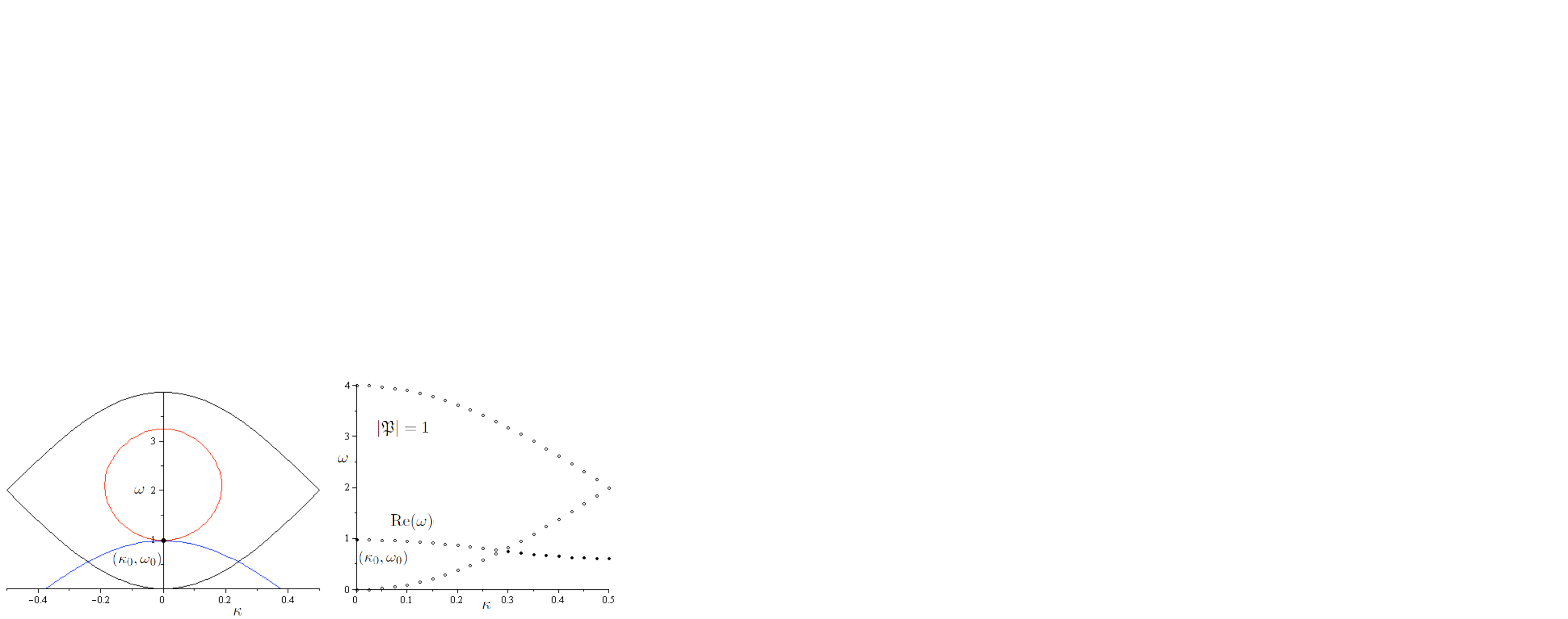}
\caption{\footnotesize{Example for a non-robust guided mode at $\kappa_0=0$ with $M_0=2$, $M_1=1$, $k_0=k_1=1$, $\gamma_0=1.029633513$, and $\gamma_1=7$. Left: The intersection point of the two relations gives the parameters $(\kappa_0=0,\omega_0\approx1)$ of the guided mode. Right: Real part of the dispersion relation in the region of one propagating harmonic.}}
 \label{Touch}
\end{figure}
 
\begin{figure}[ht] 
	\centering 
\includegraphics[width=14.7cm,height=5.8cm]{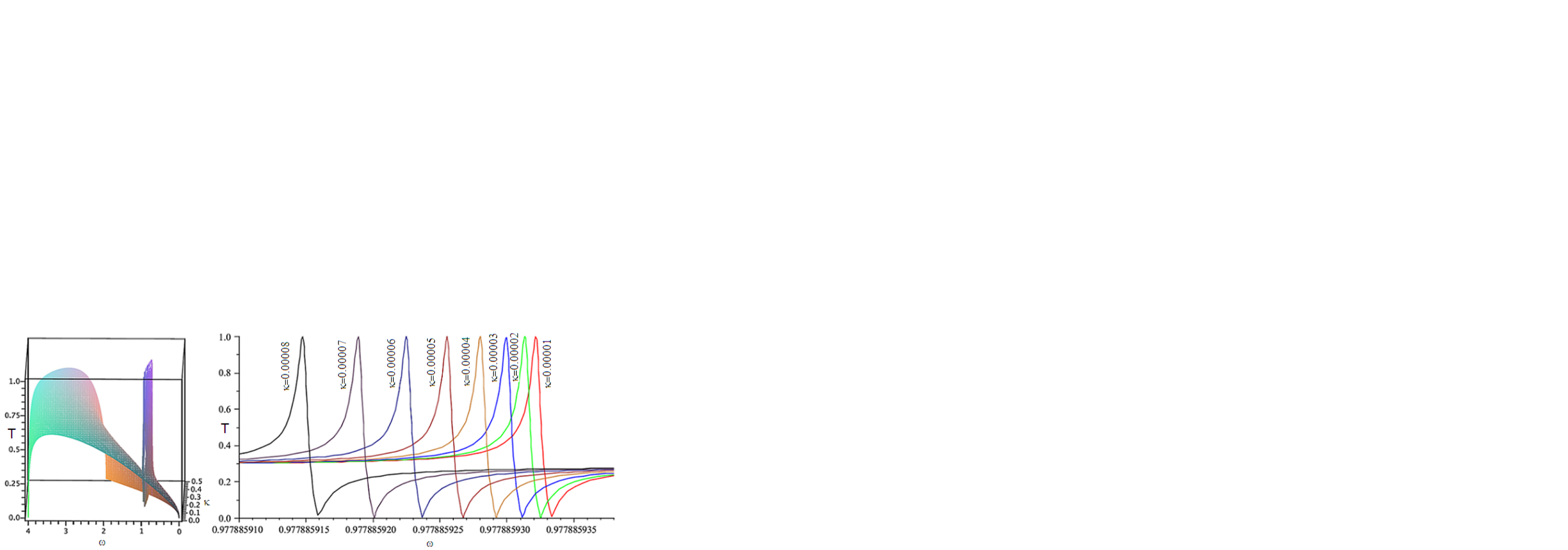}
\caption{\footnotesize{Left: The transmission coefficient for $M_0=2$, $M_1=1$, $\gamma_0=1.029633513$, $\gamma_1=7$.  {There is a guided mode at $(\kappa_0,\omega_0)=(0, 0.9778859328...)$.}  Right: with refining resolution for $\kappa$.}  Observe that, for this example, $\ell_1=0$ in the formula of Theorem~\ref{error_T}, so that both the width and the center of the resonance vary quadratically as a function of $\kappa-\kappa_0$.}
 \label{Resolution_Touch}
\end{figure}

One can construct guided modes for the system with period $N$ greater than 2.  By choosing $M_n$, $k_n$, and $\gamma_n$ to be symmetric (about some index $n$ if $N$ is odd and about a point between some $n$ and $n+1$ if $N$ is even), one can construct anti-symmetric guided modes at $\kappa=0$ in a $\kw$ region that admits only one propagating harmonic.  The idea is that, for $\kappa=0$, the system decouples into symmetric and anti-symmetric parts, and the single propagating harmonic is constant in $n$ and therefore necessarily even.  One then seeks solutions to the anti-symmetric guided-mode problem, in which propagating harmonics are automatically absent.  This idea underlies behind the existence of non-robust guided modes in open electromagnetic or acoustic waveguides \cite{ShipmanVolkov2007,Bonnet-BeStarling1994}.

In the case of period three, for example, we can make the following specific assertion.

{\theorem
\hspace{1ex}
For the coupled system of period $N=3$,
 
1.  if $\gamma_i=1$ and $k_i=1$ for $i=0,1,2$;  and $0<M_1=M_2=M<3\sqrt{21}$ and $M\ne M_0>0$, there is a guided mode at $\kappa=0$
and $\omega \in (0,3)$;

2. if $\gamma_0=\gamma_1=\gamma_2=\gamma$ and if there is a guided mode at $\kappa\!=\!0$ and $\omega\!\in\!(0,3)$ then $c_0\!=\!0$. Additionally if $M_1=M_2=M>0$, then the mode is antisymmetric about a line passing through the zeroth bead, that is $c_1=-c_2$, $a_1^-=-a_2^-$, and $b_1^+=-b_2^+$.}
\medskip

\noindent {\emph{Proof:}} 
To prove part (1), the symmetry of the structure about a line passing through the zeroth mass allows the construction of an antisymmetric guided mode at $\kappa=0$ about the same line.  The coefficients $c_i$ for the field in the waveguide satisfy
\begin{equation}
\label{seventy-eighth}
c_0=0, \quad c_1=-c_2,
\end{equation}
while the coefficients for the field in the ambient lattice satisfy
\begin{equation}
\label{seventy-ninth}
a_{0}^{-}=b_{0}^{+}=0, \quad a_{1}^{-}=-a_{2}^{-}, \quad b_{1}^{+}=-b_{2}^{+}.
\end{equation}
Taking into account these restrictions on the coefficients, the system (\ref{matrix}) with vanishing right-hand side becomes
\begin{equation}
\label{eightieth}
\begin{cases}
a_{1}^{-}=b_{1}^{+},\\
c_{1}=2i a_{1}^{-} \sin{(2\pi \theta_1)},\\
2i \sin{(2\pi \theta_1)}(\omega - \frac{3}{M})-1=0,
\end{cases}
\end{equation}
where $\sin{(2\pi \theta_1)}=i\sqrt{(\frac{5-\omega}{2})^2-1}$.
As long as $M<2\sqrt{21}$, the third of these equations determines $\omega\in(0,3/M)$.  The first two equations complete the determination of the Fourier coefficients, providing a one-parameter of family of guided modes.

To prove part (2), one sets $a_0^-=b_0^+=0$ and deduces the relations (\ref{seventy-eighth},\ref{seventy-ninth}) from  (\ref{matrix}) with vanishing right-hand side. \QED

\begin{figure}[ht]
	\centering 
\includegraphics[width=14.7cm,height=5.8cm]{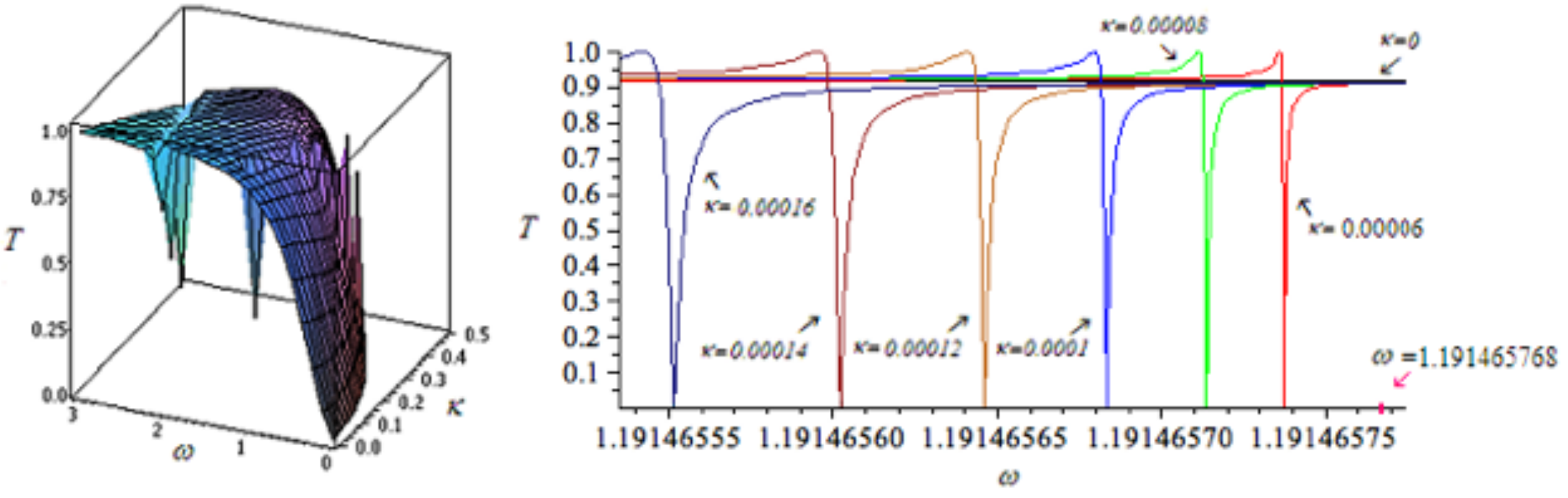}
\caption{\footnotesize{The transmission coefficient for $N=3$ with $\gamma_0=\gamma_1=\gamma_2=1$, $k_0=k_1=k_2=1$, $M_0=1$, $M_1=M_2=2$, $\kappa_0=0$, $\omega_0=1.191465768$ (left), with refining resolution for $\kappa$ (right).}}
 \label{case N=3, 1}
\end{figure}

\section{Resonant scattering}\label{sec:resonantscattering}

The analysis of resonant scattering at wavenumbers and frequencies near those of nonrobust guided modes is based on the complex-analytic connection between guided modes (for $D\kw=0$) and scattering states (for $D\kw\not=0$).  This is achieved by scaling the incident field by an eigenvalue $\ell\kw$ of $\mathbb{B}$ whose zero set coincides with the dispersion relation near a real guided-mode pair $\kwz$.  A complex perturbation analysis of the the eigenvalue, the complex transmission coefficient, and the complex reflection coefficient, all of which vanish at $\kwz$, yields asymptotic formulas for transmission anomalies.  

The analysis below follows that of \cite{ShipmanVenakides2005} for anomalous scattering by periodic dielectric slabs.  There, the Bloch wavenumber of the nonrobust guided mode vanishes $\kappa_0\!=\!0$, so that the mode is a periodic standing wave.  The consequence of this is that the transmission anomaly, to linear order in $\kappa\!-\!\kappa_0$, \emph{remains centered about the frequency $\omega_0$ of the guided mode}.  The existence of standing modes whose (minimal) period is equal to that of the structure can be proved \cite{ShipmanVolkov2007,Bonnet-BeStarling1994}, but there seems to be no proof hitherto in the literature of the existence of guided modes at nonzero~$\kappa$.  (Guided modes for $\kappa\not\!=\!0$ whose period is a multiple of that of the slab can be constructed, but these are described by a real dispersion relation and are therefore robust under perturbations of $\kappa$).

It turns out that one can show that our discrete model with period $N\!=\!2$ admits truly traveling modes ($\kappa_0\!\not=\!0$) for appropriate choices of the masses and coupling constants.  We will see that the nonvanishing of $\kappa_0$ coincides with a shifting, or detuning, of the central frequency of the resonance as $\kappa$ is perturbed from $\kappa_0$.  This detuning is linear in $\kappa\!-\!\kappa_0$, while the width of the resonance increases as $(\kappa\!-\!\kappa_0)^2$.  In section \ref{subsec:bifurcation}, we show how standing modes and traveling modes are connected through a structural parameter.  Keeping all parameters fixed except $\gamma_0$, which controls the coupling of the even-indexed sites of the waveguide to the planar lattice, we show that, if $\gamma_0$ lies above a certain critical value, the system admits no guided modes, that a standing mode ($\kappa_0\!=\!0$) is initiated at the critical coupling, and that this mode bifurcates into two guided modes traveling in opposite directions ($\kappa_0\!=\!\pm\kappa$) as $\gamma_0$ passes below the critical value.  The behavior of the transmission coefficient is complicated near the point of bifurcation, and we give an asymptotic formula for it that incorporates $\kappa$, $\omega$, and $\gamma_0$.

\subsection{Asymptotic Analysis of Transmission Near a Guided-Mode Frequency}\label{subsec:asymptotic}

Nontrivial solutions of the sourceless problem $\mathbb{B} \overrightarrow{X}=0$ occur at values of $\kappa$ and $\omega$ where the matrix $\mathbb{B}$ has a zero eigenvalue $\ell=\ell(\kappa,\omega)=0$. The relation $\ell(\kappa,\omega)=0$, or $\omega=\omega(\kappa)$ when solved for~$\omega$, is a branch of the complex dispersion relation for generalized guided modes. We analyze states that correspond to a simple zero eigenvalue $\ell$ (that is, having multiplicity $1$) occuring at a real pair $(\kappa_0,\omega_0)$ that is in the region I in Fig.~\ref{diagram_2_3} with one propagating harmonic corresponding to $\theta_0$.  By of the analyticity of $\ell\kw$ and under the generic assumption that $\partial\ell/\partial\omega\not=0$ at $\kwz$, the Weierstra{\ss} Preparation Theorem provides the following local form for the dispersion relation:
$$
\ell(\kappa,\omega)=0 \quad \Leftrightarrow \quad \omega=\omega_0+\ell_1(\kappa-\kappa_0)+\ell_2(\kappa-\kappa_0)^2+ \bigo(|\kappa-\kappa_0|^3),
$$
where $\ell_1$ is real, and $\im\,\ell_2 \ge 0$ because $\im\,\omega$ for real $\kappa$ cannot be positive due to Theorem~\ref{nonzero im}.
 
 
Since $\ell$ is of multiplicity $1$ near $\kwz$, there is an analytic change-of-basis matrix $C$  such that $\mathbb{B}$ has the form
\begin{equation}
\label{jordan}
\mathbb{B}=C\mathbb{J}C^{-1}=
C \left(
\begin{matrix}
\ell & 0\\
0 & \mathbb{\widetilde{B}}
\end{matrix}
\right)C^{-1}
\end{equation}
where the analytic matrix $\mathbb{\widetilde{B}}$ has dimension $(3N-1)\times(3N-1)$ and a bounded analytic inverse.  Let $\varphi_{mn}\kw$ be a given analytic source field, as a plane wave incident upon the waveguide from the left,
\begin{equation*}
  \varphi_{mn} = e^{2\pi i(\theta_0 m + \phi_0 n)}.
\end{equation*}
The source vector $\vec{F}$ in \eqref{matrix1} is then determined by \eqref{matrix} and this choice of source field, and $\vec{F}$ can be decomposed into its resonant and nonresonant parts
\begin{equation}\label{alpha}
  \vec{F}=\alpha C e_1+C(0,F_2),  
\end{equation}
where the complex scalar $\alpha=\alpha(\kappa,\omega)$ and the vector $F_2 \in \mathbb{C}^{3N-1}$, are analytic, and the vector $e_1\!=\!(1,0,\ldots,0)\! \in \!\mathbb{C}^{3N}$.  
We now scale this source by a constant multiple of $\ell$, $c\ell\kw\varphi_{mn}\kw$, so that it vanishes on the dispersion relation near $\kwz$ ($c\not=0$ is to be fixed later), and solve
\begin{equation}\label{BXF}
  \mathbb{B}\vec{X}=c \ell \vec{F}.
\end{equation}
The solution is
\begin{equation}\label{Xsolution}
\vec{X}=c\alpha Ce_1+c\ell C(0, \mathbb{\widetilde{B}}^{-1}F_2).
\end{equation}
The analytic vector $\vec{X}$ corresponds to a solution field $\psi_{mn}\kw$ that connects scattering states with guided modes. If $\ell(\kappa,\omega)=0$, $\psi_{mn}$ is a generalized guided mode, otherwise it is a scattering state.

For $\kw$ near $\kwz$ for which $\ell\kw\not=0$ the solution $\psi_{mn}$ in the ambient lattice satisfies the asymptotic relations
\[
\psi_{mn} \thicksim \ell e^{2 \pi i \theta_0 m}e^{2\pi i \phi_0 n}+ae^{-2 \pi i \theta_0 m}e^{2\pi i \phi_0 n} \quad  m \to -\infty,
\]
\[
\psi_{mn} \thicksim be^{2 \pi i \theta_0 m}e^{2\pi i \phi_0 n} \quad  m \to \infty.
\]
By the conservation of energy relation (\ref{conservation}), for real $(\kappa,\omega)$, we have $|\ell|^2=|a|^2+|b|^2$, which implies that $\ell$, $a$, and $b$ have a common root at $(\kappa_0,\omega_0)$. In the following analysis we use the notation $\wo=\omega-\omega_0$ and $\wk=\kappa-\kappa_0$.

The {Weierstra\ss} preparation theorem for analytic functions of two variables provides the following forms for $\ell$, $a$, and $b$\,:
\begin{equation} \nonumber
\begin{array}{l}
\ell\kw=e^{i\rho_1}[\wo+\ell_1\wk+\ell_2 \wk^2+\bigo(|\wk|^3)][1+\bigo(|\wk|+|\wo|)],\\
a\kw=e^{i\rho_2}[\wo+r_1\wk+r_2\wk^2+\bigo(|\wk|^3)][r_0+\bigo(|\wk|+|\wo|)],\\
b\kw=e^{i\rho_3}[\wo+t_1\wk+t_2\wk^2+\bigo(|\wk|^3)][t_0+\bigo(|\wk|+|\wo|)],
\end{array}
\end{equation}
in which $r_0$, and $t_0$ are positive real numbers and we have chosen $c$ so that the corresponding coefficient for $\ell$ is unity.
%
%
Using these expressions, we expand  
%
%
the relation $|\ell|^2=|a|^2+|b|^2$ for real $(\wk,\wo)$ to obtain the following relations among the coefficients:
\begin{equation}
\begin{array}{ll}
1=r_0^2+t_0^2 \quad & (\text{$\wo^2$ term}),\\
\ell_1^2=r_0^2 |r_1|^2+t_0^2 |t_1|^2 \quad & (\text{$\wk^2$ term}),\\
\ell_1=r_0^2 \re(r_1)+t_0^2 \re(t_1) \quad & (\text{$\wo \wk$ term}),\\
\re(\ell_2)=r_0^2 \re(r_2)+t_0^2 \re(t_2) \quad & (\text{$\wo \wk^2$ term}),\\
\ell_1 \re(\ell_2)=r_0^2 \re(r_2\bar{r}_1)+t_0^2 \re(t_2 \bar{t}_1) \quad &(\text{$\wk^3$ term}),\\
|\ell_2|^2\!+\!2\ell_1\re(\ell_3)\!=\!r_0^2[|r_2|^2\!+\!2\re(r_1\bar{r}_3)]\!+\!t_0^2[|t_2|^2\!+\!2\re(t_1\bar{t}_3)]  & (\text{$\wk^4$ term}).
\end{array}
\end{equation}
Because of Theorem~\ref{nonzero im}, which says that $\im\,\omega\leq 0$ if $\kappa$ is real and $\ell\kw=0$, we find that $\ell_1$ must be real-valued and that $\im\,\ell_2\geq0$.
Because of the equations $r_0^2+t_0^2=1$ and $\ell_1=r_0^2\re(r_1)+t_0^2\re(t_1)$, $\ell_1$ lies between $\re(r_1)$ and $\re(t_1)$. 


{\theorem \label{thm:linearcoefficients} $\ell_1=t_1=r_1\in\R$.}

\medskip
\noindent {\emph{Proof:}} Suppose $r_1=r_{1R}+ir_{1I}$ and $t_1=t_{1R}+it_{1I}$, then it follows that
$$
\ell_1=r_0^2 r_{1R}+t_0^2 t_{1R} \quad \text{and} \quad \ell_1^2=r_0^2 r_{1R}^2+t_0^2t_{1R}^2+r_0^2r_{1I}^2+t_0^2t_{1I}^2.
$$
By convexity, the first equality implies
$$
\ell_1^2 \le r_0^2r_{1R}^2+t_0^2t_{1R}^2.
$$
This is consistent with the second equality if and only if $r_{1I}=t_{1I}=0$ and $r_{1R}=t_{1R}$, which yields $\ell_1=r_1=t_1$.
\QED
\medskip

\begin{figure}
	\centering 
\includegraphics[width=9.3cm,height=10.5cm]{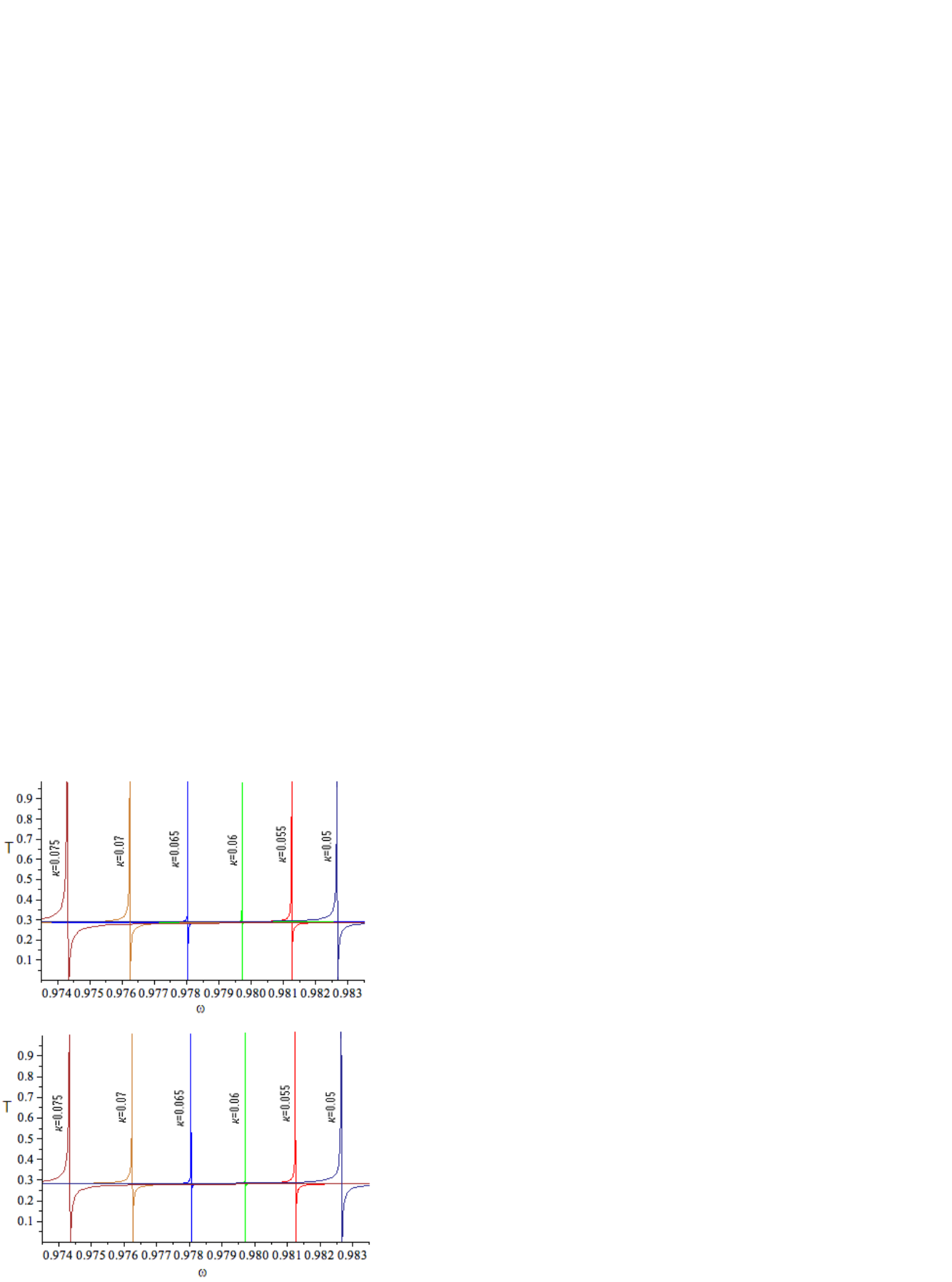}
\caption{\footnotesize{The transmission coefficient for $M_0=2$, $M_1=1$, $\gamma_0=1$, $\gamma_1=7$. Upper: The exact formula for the transmission coefficient. Lower: Approximation with second order term in $\kappa$, $\eta \approx 0.767728$, $t_0 \approx 0.3142988$, $r_0\approx 0.94932$.
{A guided mode is supported at the parameters $(\kappa_0,\omega_0)\approx(0.0616, 0.9792)$.}}}
 \label{GM_Orig_Approx}
\end{figure}

We show now how to obtain a formula that approximates the transmission anomalies. According to the above theorem we use the expansions for $a$ and $b$ including terms of the second order in $\wk$, that is
\begin{equation}
\label{forms}
\begin{array}{l}
 \ell=e^{i\rho_1}(\wo+\ell_1\wk+\ell_2\wk^2+\ldots)(1+c_1\wo+c_2\wk+\ldots),\\
  a= r_0e^{i\rho_2}(\wo+\ell_1\wk+t_2\wk^2+\ldots)(1+p_1\wo+p_2\wk+\ldots),\\
  b= t_0e^{i\rho_3}(\wo+\ell_1\wk+r_2\wk^2+\ldots)(1+q_1\wo+q_2\wk+\ldots).
  \end{array}
\end{equation}
In the first factors, the higher-order terms are $\bigo(|\wk|^3)$, in the second, they are $\bigo(\wk^2+\wo^2)$. The transmission cefficient $T$ depends on the absolute value of the ratio $b/a$,
\begin{equation}
\label{transmission}
T=\frac{|b|}{|\ell|}=\frac{|b|}{\sqrt{|a|^2+|b|^2}}=\frac{|b/a|}{\sqrt{1+|b/a|^2}},
\end{equation} 
and $b/a$ has form
\begin{equation}
\frac{b}{a}=e^{i \rho}\frac{t_0}{r_0} \frac{(\wo+t_1\wk+t_2\wk^2+\bigo(|\wk|^3))}{(\wo+r_1\wk+r_2\wk^2+\bigo(|\wk|^3))}(1+\eta_1\wo+\eta_2\wk+\bigo(|\wk|+|\wo|)),
\end{equation} 
in which $\rho=\rho_3-\rho_2$, $\eta_1=q_1-p_1$, $\eta_2=q_2-p_2$.
%
%
The approximation
\begin{equation}
\Bigl|\frac{b}{a}\Bigr| \approx \frac{t_0|\wo+t_1\wk+t_2\wk^2|}{r_0|\wo+r_1\wk+r_2\wk^2|}|1+\eta\wo|
\end{equation}
yields the following approximation for the transmission coefficient
%
%
\begin{equation}
\label{trans_app}
T^2\approx \frac{t_0^2|\wo+t_1\wk+t_2\wk^2|^2|1+\eta \wo|^2}{{r_0^2|\wo+r_1\wk+r_2\wk^2|^2+t_0^2|\wo+t_1\wk+t_2\wk^2|^2|1+\eta\wo|^2}},
\end{equation}
which agrees well with the exact formula (see Fig.~\ref{GM_Orig_Approx}).  One can see on those graphs that a sharp resonance emanates from the guided-mode frequency $\omega_0$ as the wave number $\kappa$ is perturbed from~$\kappa_0$. The anomaly widens quadratically as a function of $\wk$ and it is detuned linearly away from the guided mode frequency $\omega_0$, which indicates that $\ell_1\not=0$.
This is formula generalizes that of \cite{ShipmanVenakides2005}, where it was assumed that $\ell_1=0$ because the structure was symmetric with respect to a line perpendicular to the waveguide and the guided mode was a standing wave ($\kappa_0=0$).

One can prove that one commits an error of $\bigo(|\wk|+\wo^2)$ in the approximation \eqref{trans_app}.

{\theorem \label{error_T} If $\ell(\kappa,\omega)$ has a root at $(\kappa_0,\omega_0)\in \mathbb{R}^2$; the partial derivatives of $\ell$, $a$, and $b$ with respect to $\omega$ do not vanish at  $(\kappa_0,\omega_0)$; and $\im\,\ell_2\ne0$ in the form $(\ref{forms})$, then the error in the approximation \eqref{trans_app} is of order $\bigo(|\wk|+\wo^2)$ and the following approximation holds:
\begin{equation}\label{trans_app2}
T(\kappa,\omega)=t_0\frac{|\wo+\ell_1\wk+t_2\wk^2|}{|\wo+\ell_1\wk+\ell_2\wk^2|}|1+\zeta_1\wo|+\bigo(|\wk|+\wo^2)
\end{equation}
as $(\wk,\wo) \to(0,0)$ in $\mathbb{R}^2$, where $\zeta_1=q_1-c_1$. }

\medskip
\noindent {\emph{Proof:}} 
We shall prove only the formula \eqref{trans_app2}; the error in \eqref{trans_app} can be proved similarly.
%
\begin{equation} \label{nashaT}
\begin{array}{l}
T=\displaystyle\displaystyle \left| \frac{b(\kappa,\omega)}{\ell(\kappa,\omega)} \right|=t_0\left|\frac{(\wo+\ell_1\wk+t_2\wk^2+t_3\wk^3+\cdots)}{(\wo+\ell_1\wk+\ell_2\wk^2+\ell_3\wk^3+\cdots)}(1+\zeta_1\wo+\zeta_2\wk+\cdots)\right|. \\
\end{array}
\end{equation}
Since $\ell_1$ is real-valued and $\im\,\ell_2\not=0$, the denominator can be written as
\begin{equation}
{\wo\!+\!\ell_1\wk\!+\!\ell_2\wk^2\!+\!\ell_3\wk^3\!+\!\bigo(|\wk|^{4})}\!=\!\left({\wo\!+\!\ell_1\wk\!+\!\ell_2\wk^2}\right)\left({1\!+\!\displaystyle \frac{\ell_3\wk^3\!+\!\bigo(|\wk|^{4})}{\wo\!+\!\ell_1\wk\!+\!\ell_2\wk^2}}\right).
\end{equation}
Denote $\epsilon=\displaystyle \frac{\ell_3\wk^3+\bigo(|\wk|^{4})}{\wo+\ell_1\wk+\ell_2\wk^2}=\bigo(|\wk|)$.  Using $|\epsilon|<1$ we obtain
\begin{equation}
\renewcommand{\arraystretch}{2.5}
\begin{array}{l}
\displaystyle\frac{(\wo+\ell_1\wk+t_2\wk^2+t_3\wk^3+\cdots)}{(\wo+\ell_1\wk+\ell_2\wk^2+\ell_3\wk^3+\cdots)}
=\frac{(\wo+\ell_1\wk+t_2\wk^2+t_3\wk^3+\cdots)}{\wo+\ell_1\wk+\ell_2\wk^2} \cdot (1-\epsilon+\epsilon^2-\cdots)\\
=\displaystyle \frac{(\wo+\ell_1\wk+t_2\wk^2)}{(\wo+\ell_1\wk+\ell_2\wk^2)}(1-\epsilon+\cdots)+\frac{(\sum_{j=3}^\infty t_j\wk^j)}{(\wo+\ell_1\wk+\ell_2\wk^2)}(1-\epsilon+\cdots).
\end{array}
\end{equation}
Thus the expression in absolute values on the right of $(\ref{nashaT})$ is
\begin{equation}
\begin{array}{l}
\displaystyle \frac{(\wo+\ell_1\wk+t_2\wk^2)}{(\wo+\ell_1\wk+\ell_2\wk^2)}(1-\epsilon+\cdots)(1+\zeta_1\wo+\zeta_2\wk+\cdots)\\
\vspace{-1.5ex} \\
\displaystyle +\frac{(\sum_{j=3}^{\infty}t_j\wk^j)}{(\wo+\ell_1\wk+\ell_2\wk^2)}(1-\epsilon+\cdots)(1+\zeta_1\wo+\zeta_2\wk+\cdots)
\end{array}
\end{equation}
Again, because $\im\,\ell_2\not=0$, the second term is $\bigo(|\wk|)$, and we obtain the result.

\subsection{Bifurcation of Guided Modes and Resonance}\label{subsec:bifurcation}
In this section, will see how the strength of the coupling between the waveguide and the ambient lattice acts as a tangent bifurcation parameter for the creation and splitting of a guided mode.  We will study the simplest case of period 2 with $k_0=k_1=k$, in which we fix all parameters except one of the coupling constants.  When this constant is lowered to a specific value, a guided mode is created at $\kappa=\kappa_0=0$ (Fig.~\ref{Cross}) and is thus a standing wave exponentially confined to the waveguide.  When the constant is lowered further, the mode splits into two guided modes at $\kappa=\pm\kappa_0\not=0$ (Fig.~\ref{Touch}) traveling in opposite directions along the waveguide.

Such a tangent bifurcation connects the case of $\ell_1=0$ in the transmission coefficient in Theorem \ref{error_T}, to the case of $\ell_1 \ne 0$.  Indeed, for a standing mode ($\kappa_0=0$), $\wk=\kappa$ in the theorem and the symmetry of the transmission coefficient with respect to $\kappa$ implies that $\ell_1=0$.  In this case, the transmission formula shows that both the center of the anomaly and the distance between the peak and dip vary quadratically in $\wk$.  On the other hand, when $\kappa_0\not=0$ and the mode is traveling, we typically have $\ell_1\not=0$ and thus there is a detuning of the central frequency of the anomaly away from the frequency $\omega_0$ of the guided mode; the detuning is related linearly to $\wk$ (to leading order), while the anomaly widens still only at a quadratic rate.

We will need the following technical lemma.

{\lemma \label{real_functions} Suppose that, for $N=2$ and for fixed real values of $M_0$, $M_1$, $k_0=k_1=k$, $\gamma_0$ and~$\gamma_1$, there is a unique real pair $(\kappa_0,\omega_0)$ in an open set $U$ of the real $(\kappa,\omega)$-region $I$ (Fig.~\ref{diagram_2_3}) of one propagating harmonic that admits a true guided mode, that is $\ell(\kappa_0,\omega_0)=0$. Assume in addition the generic conditions $\im\,\ell_2\not=0$, $\frac{\partial \ell}{\partial \omega}(\kappa_0,\omega_0) \ne 0$, $\frac{\partial a}{\partial \omega}(\kappa_0,\omega_0) \ne 0$, $\frac{\partial b}{\partial \omega}(\kappa_0,\omega_0) \ne 0$ hold.  
\begin{enumerate}
\item There exist intervals $I$ about $\kappa_0$ and $V$ about $\omega_0$ and analytic real-valued functions $\omega_a,\omega_b:I\to V$ such that $a(\kappa,\omega_a(\kappa))=0$, $b(\kappa,\omega_b(\kappa))=0$ and $\omega_a(\kappa_0)=\omega_b(\kappa_0)=\omega_0$.  Thus $\omega_a(\kappa)$, $\omega_b(\kappa)$ for $\kappa \in I \backslash\{\kappa_0\}$ describe real frequencies for which the transmission $T$ reaches presicely $100\%$ (peak) and $0\%$ (dip), respectively.
\item Either $\omega_a(\kappa)>\omega_b(\kappa)$, for all $\kappa \in I\backslash\{\kappa_0\}$, which means the peak in the transmission comes to the right of the dip, or $\omega_a(\kappa)<\omega_b(\kappa)$, $\kappa\in I \backslash \{\kappa_0\}$, which implies the peak in the transmission comes to the left of the dip.
\end{enumerate} 
}

\noindent {\emph{Proof:}} According to $(\ref{matrix})$ the zero-sets for $a(\kappa,\omega)$ and $b(\kappa,\omega)$ are defined by
\begin{equation}
\det \left(
\begin{array}{cccc}
0 & -(\gamma_0+\gamma_1) & 0 & -2\gamma_0 \\
 0& -\gamma_1  & -2i\sin(2\pi \theta_1) & 0 \\
 -\gamma_0 & \omega-\frac{2k}{M_0}+\frac{2k\cos(\pi\kappa)}{\sqrt{M_0M_1}}&-\gamma_0 & 2 \omega -\frac{4k}{M_0}\\
 -\gamma_1& \omega-\frac{2k}{M_1} +\frac{2k\cos(\pi\kappa)}{\sqrt{M_0M_1}} &\gamma_1 & \frac{4k\cos(\pi\kappa)}{\sqrt{M_0M_1}}
\end{array} 
\right)=0,
\end{equation}

\begin{equation}
\det \left(
\begin{array}{ccc}
\gamma_1-\gamma_0 &4i\sin(2\pi\theta_1)& -2\gamma_0 \\
\omega-\frac{2k}{M_0}+\frac{2k\cos(\pi\kappa)}{\sqrt{M_0M_1}} & -\gamma_0 & 2\omega-\frac{4k}{M_0} \\
\omega-\frac{2k}{M_1}+\frac{2k\cos(\pi\kappa)}{\sqrt{M_0M_1}} & \gamma_1 & \frac{4k\cos(\pi\kappa)}{\sqrt{M_0M_1}}
\end{array}
\right)=0,
\end{equation}
respectively, which are real-valued functions of $\kw$ in region $I$ of the real $\kw$ plane, with $\sin(2\pi\theta_1)=i\sqrt{(2-\frac{\omega}{2}+\cos(\pi\kappa))^2-1}$. 
Both of these conditions are satisfied at $\kwz$.
Because of the conditions $\frac{\partial a}{\partial \omega}|_{(\kappa_0,\omega_0)}\ne0$ and  $\frac{\partial b}{\partial \omega}|_{(\kappa_0,\omega_0)}\ne0$, part (1) of the theorem follows from the implicit function theorem.  These functions have expansions with real coefficients,
\begin{eqnarray}
&& \omega_a(\kappa)=\omega_0-\ell_1(\kappa-\kappa_0)-r_2(\kappa-\kappa_0)^2-\ldots,\quad (a=0) \label{a}
\\
&& \omega_b(\kappa)=\omega_0-\ell_1(\kappa-\kappa_0)-t_2(\kappa-\kappa_0)^2-\ldots,\quad (b=0)  \label{b}
\end{eqnarray} 
in which the coefficients of the linear terms are equal by Theorem~\ref{thm:linearcoefficients}.

In \cite[Theorem 20(4)]{Shipman2010}, it is proved that the assumption $\im\,\ell_2\not=0$ implies that $r_2\not=t_2$, from which part (2) follows.
\QED

\begin{figure} 
	\centering 
\includegraphics[width=16cm,height=4.5cm]{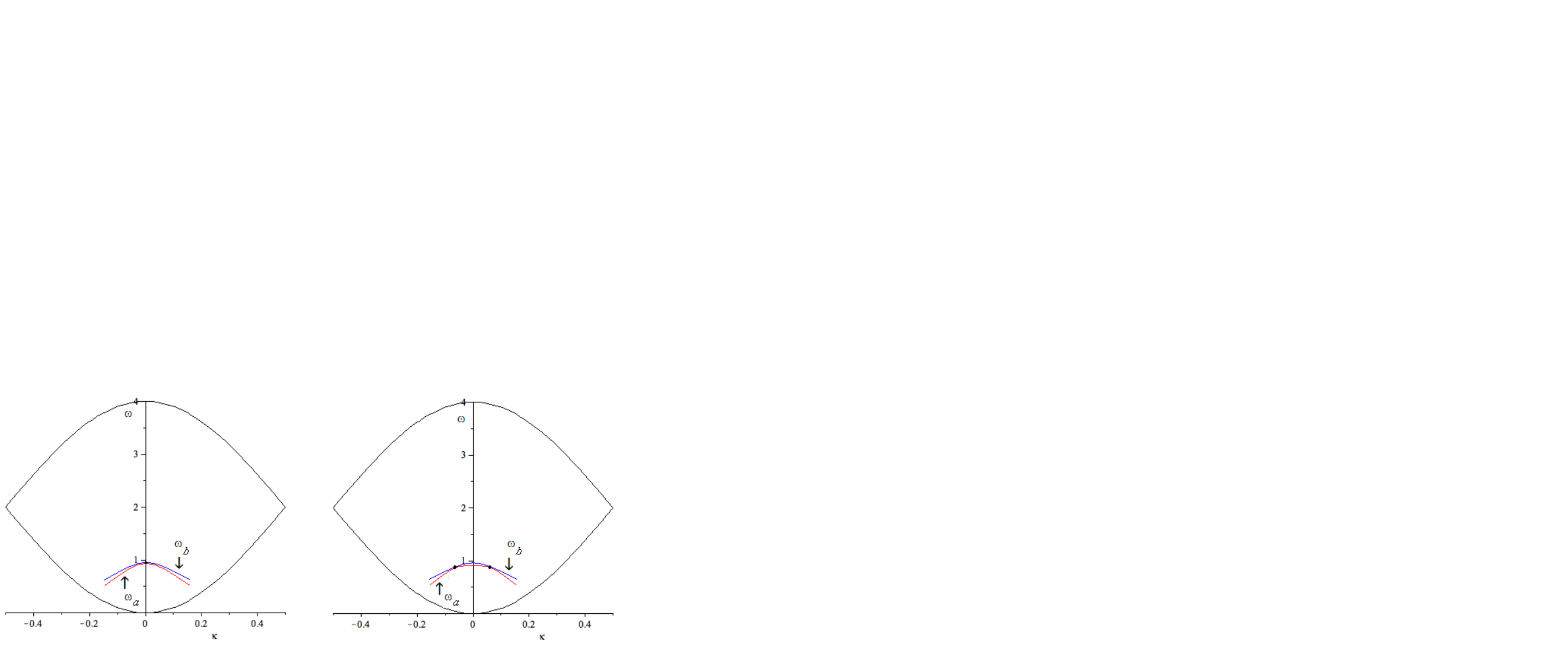}
\caption{\footnotesize{Schematic of the functions $\omega_a$ and $\omega_b$ for two cases $M_0=2$, $M_1=1$, $k=1$, $\gamma_1=7$. The solid dot represents where the guided modes exist. Left: At $\kappa_0^*=0$ for $\gamma_0=\gamma_0^*$. Right: At $\kappa_0=\mp 0.0616$ for $\gamma_0=1$.}}
 \label{frequencies}
\end{figure}

The analysis  of the transmission anomaly relies on the following conditions:
\begin{equation}
\begin{array}{c}
|\ell(\kappa,\omega,\gamma_0)|^2=|a(\kappa,\omega,\gamma_0)|^2+|b(\kappa,\omega,\gamma_0)|^2 \quad \text{for $\kappa,  \omega,\gamma_0 \in \mathbb{R}$}\\
\text{ if $\ell(\kappa,\omega,\gamma_0)=0$ for $\kappa \in \mathbb{R}$, then $\im(\omega)\le0$.}  
\end{array}
\end{equation} 
\begin{equation}
\label{simultaneously}
\ell(\kappa_0^*,\omega_0^*,\gamma_0^*)=0, \quad a(\kappa_0^*,\omega_0^*,\gamma_0^*)=0, \quad b(\kappa_0^*,\omega_0^*,\gamma_0^*)=0
\end{equation} 
where $(\kappa_0^*=0,\omega_0^*,\gamma_0^*) \in \mathbb{R}^3$ is the bifurcation point.

The following conditions hold generically:
\begin{equation}
\label{generic_bifur}
\begin{array}{ccc}
\displaystyle \frac{\partial \ell}{\partial \omega}(\kappa_0^*,\omega_0^*,\gamma_0^*)\ne0, & \displaystyle \frac{\partial a}{\partial \omega}(\kappa_0^*,\omega_0^*,\gamma_0^*)\ne0,  & \displaystyle \frac{\partial b}{\partial \omega}(\kappa_0^*,\omega_0^*,\gamma_0^*)\ne0.
\end{array}
\end{equation} 
The curves $a(\kappa,\omega,\gamma_0)=0$ and $b(\kappa,\omega,\gamma_0)=0$ for real values of $\kappa$ near the bifurcation point describe frequencies $\omega_a$, $\omega_b$ of the reflected and transmitted coefficients, respectively, which correspond peaks  and dips  of the transmission.

{\theorem \label{thm:bifurcation} Suppose that for the period $N=2$ system with fixed real values of $M_0$, $M_1$, $k_1=k_2=k$, and $\gamma_1$, there exists a unique triple $(\kappa_0^*=0,\omega_0^*,\gamma_0^*) \in \mathbb{R}^3$ with $(\omega_0^*,\gamma_0^*)$ in the regime of one propagating harmonic, such that $\ell(\kappa_0^*,\omega_0^*,\gamma_0^*)=0$.  Let $\ell(\kappa,\omega,\gamma_0)=\mathcal{L}_1(\kappa,\omega,\gamma_0)+i\mathcal{L}_2(\kappa,\omega,\gamma_0)$, where $\mathcal{L}_1=\re(\ell)$, $\mathcal{L}_2=\im(\ell)$ and $\mathcal{L}_1$, $\mathcal{L}_2$ are real-analytic functions of the real triple $(\kappa,\omega,\gamma_0)$. Assume $(\ref{generic_bifur})$ hold and
\begin{equation}\label{L}
\det \left(
\begin{array}{cc}
\frac{\partial \mathcal{L}_1}{\partial \omega}(\kappa_0^*,\omega_0^*,\gamma_0^*) & \frac{\partial \mathcal{L}_1}{\partial \gamma_0}(\kappa_0^*,\omega_0^*,\gamma_0^*) \\
\frac{\partial \mathcal{L}_2}{\partial \omega}(\kappa_0^*,\omega_0^*,\gamma_0^*) & \frac{\partial \mathcal{L}_2}{\partial \gamma_0}(\kappa_0^*,\omega_0^*,\gamma_0^*)
\end{array}
\right) \ne 0.
\end{equation}
Then there are intervals $I$ about $\kappa_0^*$, $J$ about $\gamma_0^*$, and $V$ about $\omega_0^*$ and smooth real-valued functions $\omega_a, \omega_b:I \times J \to V$, $g:I \to J$, $W:I \to V$ such that
\begin{eqnarray*}
   &&a(\kappa,\omega_a(\kappa,\gamma_0),\gamma_0)=0 \text{ and }\,  b(\kappa,\omega_b(\kappa,\gamma_0),\gamma_0)=0 \text{ for all }\, (\kappa,\gamma_0)\in I\times J; \\
   && \ell(\kappa_0,W(\kappa_0),g(\kappa_0))=0 \text{ for all }\, \kappa_0\in I , \\
   && g(\kappa_0^*)=\gamma_0^* \text{ and }\,  W(\kappa_0^*)= \omega_0^*.    
\end{eqnarray*}
Let us make the generic assumption that $g''(\kappa_0^*)<0$ (resp. $g''(\kappa_0^*)>0$) and that $\omega_a(\kappa,\gamma_0^*)<\omega_b(\kappa,\gamma_0^*)$ for some $\kappa\in I$ (an analogous conclusion holds for ``\,$>$\!").

The system undergoes a bifurcation at $\gamma_0=\gamma_0^*$:
\begin{enumerate}
\item For $\gamma_0=\gamma_0^*$, there is a unique $\kappa_0\in I$ such that $g(\kappa_0)=\gamma_0$, namely, $\kappa_0=\kappa_0^*=0$.  Moreover, $W(\kappa_0^*)=\omega_a(\kappa_0^*,\gamma_0^*)=\omega_b(\kappa_0^*,\gamma_0^*)=\omega_0^*$; and $\omega_a(\kappa,\gamma_0^*)< \omega_b(\kappa,\gamma_0^*)$ for all $\kappa \in I\backslash\{\kappa_0^*\}$.
%
\item For each $\gamma_0\in J$ with $\gamma_0<\gamma_0^*$ (resp. $\gamma_0>\gamma_0^*$), there exists exactly one $\kappa_0>0$ in $I$ such that $\gamma_0=g(\pm\kappa_0)$, $\omega_a(\pm\kappa_0,\gamma_0)=\omega_b(\pm\kappa_0,\gamma_0)=W(\pm\kappa_0)$; and $\omega_a(\kappa,\gamma_0)<\omega_b(\kappa,\gamma_0)$ for all $\kappa \in I\backslash\{-\kappa_0,\kappa_0\}$.
%
\item For each $\gamma_0\in J$ with $\gamma_0>\gamma_0^*$ (resp. $\gamma_0<\gamma_0^*$), there exists no $\kappa_0$ in $I$ such that $\gamma_0=g(\kappa_0)$; and $\omega_a(\kappa,\gamma_0)<\omega_b(\kappa,\gamma_0)$  for all $\kappa \in I$.
\end{enumerate}
}
\noindent {\emph{Proof:}}
The existence of the stated intervals and real-analytic functions $\omega_{a,b}$, $g$, and $W$ is a consequence of the implicit function theorem.  Because of the symmetry of $\ell$, $a$, and $b$ in $\kappa$, both $g$ and $W$ are also symmetric.  This and the nonvanishing of the second derivative $g''(\kappa_0^*)$ give rise to the three cases depending on $\gamma_0$.  The equality $\omega_a(\kappa_0,g(\kappa_0))=\omega_b(\kappa_0,g(\kappa_0))=W(\kappa_0)$ comes from the conservation of energy relation $|a|^2 + |b|^2 = |\ell|^2$ for real $\kw$.  Lemma~\ref{real_functions} guarantees that $\omega_a(\kappa,\gamma_0)\leq\omega_b(\kappa,\gamma_0)$  for all $\kappa \in I$ or  $\omega_a(\kappa,\gamma_0)\geq\omega_b(\kappa,\gamma_0)$  for all $\kappa \in I$.
\QED
\smallskip


The transmission coefficient near the bifurcation point depends delicately on the three analytic parameters $\omega$, $\kappa$, and $\gamma_0$.  To obtain an asymptotic formula for the transmission anomaly near the bifurcation, we use, in place of $\gamma_0$, the wavenumber $\kappa_0$ of a guided mode at an isolated pair $\kwz$ in real $\kw$-space; indeed, Theorem~\ref{thm:bifurcation} tells us that $\gamma_0$ is an analytic function of $\kappa_0$.  We do a complex-analytic perturbation analysis in the variables $(\wk=\kappa-\kappa_0,\wo=\omega-\omega_0,\kappa_0)$ about $(0,0,0)$, keeping in mind that $\omega_0=W(\kappa_0)$ depends analytically on $\kappa_0$.  The {Weierstra\ss}  preparation theorem for analytic functions of three variables  provides the following expansions for $\ell$, $a$, and $b$ near $(\wk,\wo,\kappa_0)=(0,0,0)$:
\begin{align}
\ell= e^{i\psi_1}[\widetilde{\omega}+\ell_{1,0}\kappa_0+\ell_{0,1}\widetilde{\kappa}+ \ell_{1,1}\kappa_0 \widetilde{\kappa} +\ldots+\ell_{i,j}{\kappa_0}^{i}\widetilde{\kappa}^j+\ldots] \nonumber \\
\quad {} \times [\lambda_0+\lambda_1\widetilde{\omega}+\lambda_2\kappa_0+\lambda_3\widetilde{\kappa}+\ldots] \label{expan_l} \\
a= e^{i\psi_2}[\widetilde{\omega}+r_{1,0}\kappa_0+r_{0,1}\widetilde{\kappa}+r_{1,1}\kappa_0\widetilde{\kappa}+\ldots+r_{i,j}{\kappa_0}^i{\widetilde{\kappa}}^j+\ldots] \nonumber \\
\quad {} \times [\rho_0+\rho_1\widetilde{\omega}+\rho_2\kappa_0+\rho_3\widetilde{\kappa}+\ldots]   \label{expan_a=0} \\
b= e^{i\psi_3}[\widetilde{\omega}+t_{1,0}\kappa_0+t_{0,1}\widetilde{\kappa}+t_{1,1}\kappa_0\widetilde{\kappa}+\ldots+t_{i,j}{\kappa_0}^i {\widetilde{\kappa}}^j+\ldots] \nonumber \\
\quad {} \times [\tau_0+\tau_1\widetilde{\omega}+\tau_2\kappa_0+\tau_3\widetilde{\kappa}+\ldots]    \label{expan_b=0}
\end{align}  
Taking into account the symmetry of these functions in $\wk$ and $\kappa_0$,
we obtain
\begin{eqnarray}
\ell = e^{i\psi_1}[\widetilde{\omega}+\ell_{1,1}\kappa_0\widetilde{\kappa}+\ell_{0,2}{\widetilde{\kappa}}^2+\ldots] [1+L_1\widetilde{\omega}+L_2\kappa_0\widetilde{\kappa}+L_3{\widetilde{\kappa}}^2+\ldots]  \label{expan_l_new} \\
a = e^{i\psi_2}\rho_0 [\widetilde{\omega}+r_{1,1}\kappa_0\widetilde{\kappa}+r_{0,2}{\widetilde{\kappa}}^2+\ldots] [1+P_1\widetilde{\omega}+P_2{\kappa_0}\widetilde{\kappa}+P_3{\widetilde{\kappa}}^2+\ldots]   \label{expan_a_new} \\
b = e^{i\psi_3} \tau_0 [\widetilde{\omega}+t_{1,1}\kappa_0\widetilde{\kappa}+t_{0,2}{\widetilde{\kappa}}^2+\ldots] [1+Q_1\widetilde{\omega}+Q_2{\kappa_0}\widetilde{\kappa}+Q_3{\widetilde{\kappa}}^2+\ldots]    \label{expan_b_new=0}
\end{eqnarray} 
Inserting these expressions into the law of conservation of energy for real $(\wk,\wo,\kappa_0)$  and matching like terms yields relations among the coefficients; for example,
\begin{equation}
\begin{array}{lcl}
(\text{$\wo^2$ term}) & &1={\rho_0}^2+{\tau_0}^2\\
(\text{$\wo\kappa_0 \wk$ term}) & &\re(\ell_{1,1})=\rho_0^2 \re(r_{1,1})+\tau_0^2 \re(t_{1,1}), \\
(\text{${\kappa_0}^2\wk^2$ term}) & &|\ell_{1,1}|^2={\rho_0}^2 |r_{1,1}|^2+{\tau_0}^2|t_{1,1}|^2,\\
(\text{$\wo \wk^2$ term}) & &\re(\ell_{0,2})=\rho_0^2 \re(r_{0,2})+\tau_0^2 \re(t_{0,2}), \\
(\text{${\wk}^4$ term}) & &|\ell_{0,2}|^2=\rho_0^2 |r_{0,2}|^2+\tau_0^2 |t_{0,2}|^2.\\
\end{array}
\end{equation}

\begin{figure}[t]
	\centering 
\includegraphics[width=15.0cm,height=9.5cm]{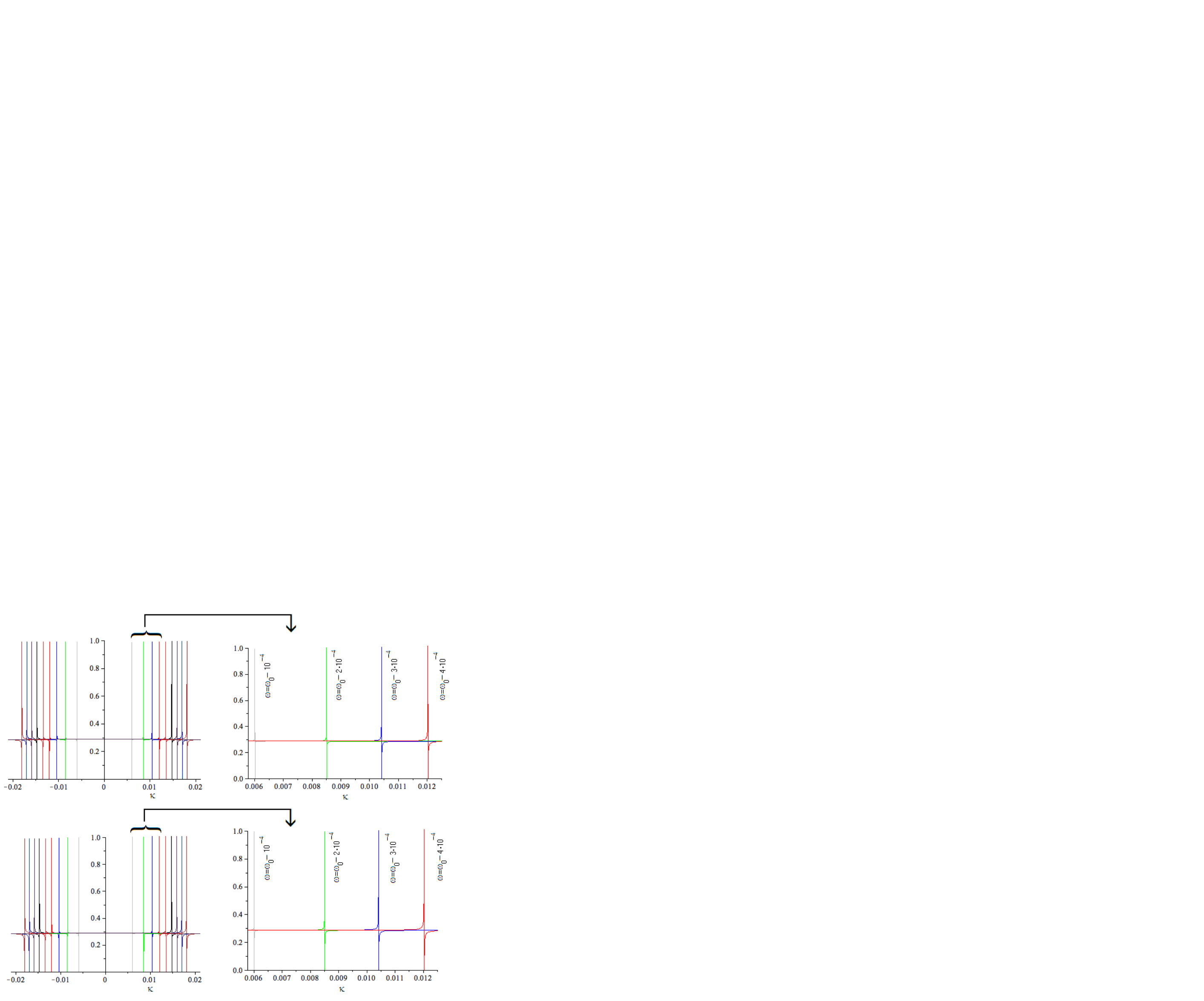}
\caption{\footnotesize{At bifurcation: The transmission coefficient for $M_0=2$, $M_1=1$, $k_1=k_2=1$, $\gamma_0=\gamma_0^*=1.029633513$, $\gamma_1=7$ near the parameters of a guided mode, $\kappa_0^*=0$, $\omega_0^*=0.9778859328$. Upper: Direct calculation of the transmission coefficient. Lower: The asymptotic formula for $|b|/|\ell|$.}}
 \label{bifur_itself}
\end{figure}

\begin{figure}
	\centering 
\includegraphics[width=15.0cm,height=8.5cm]{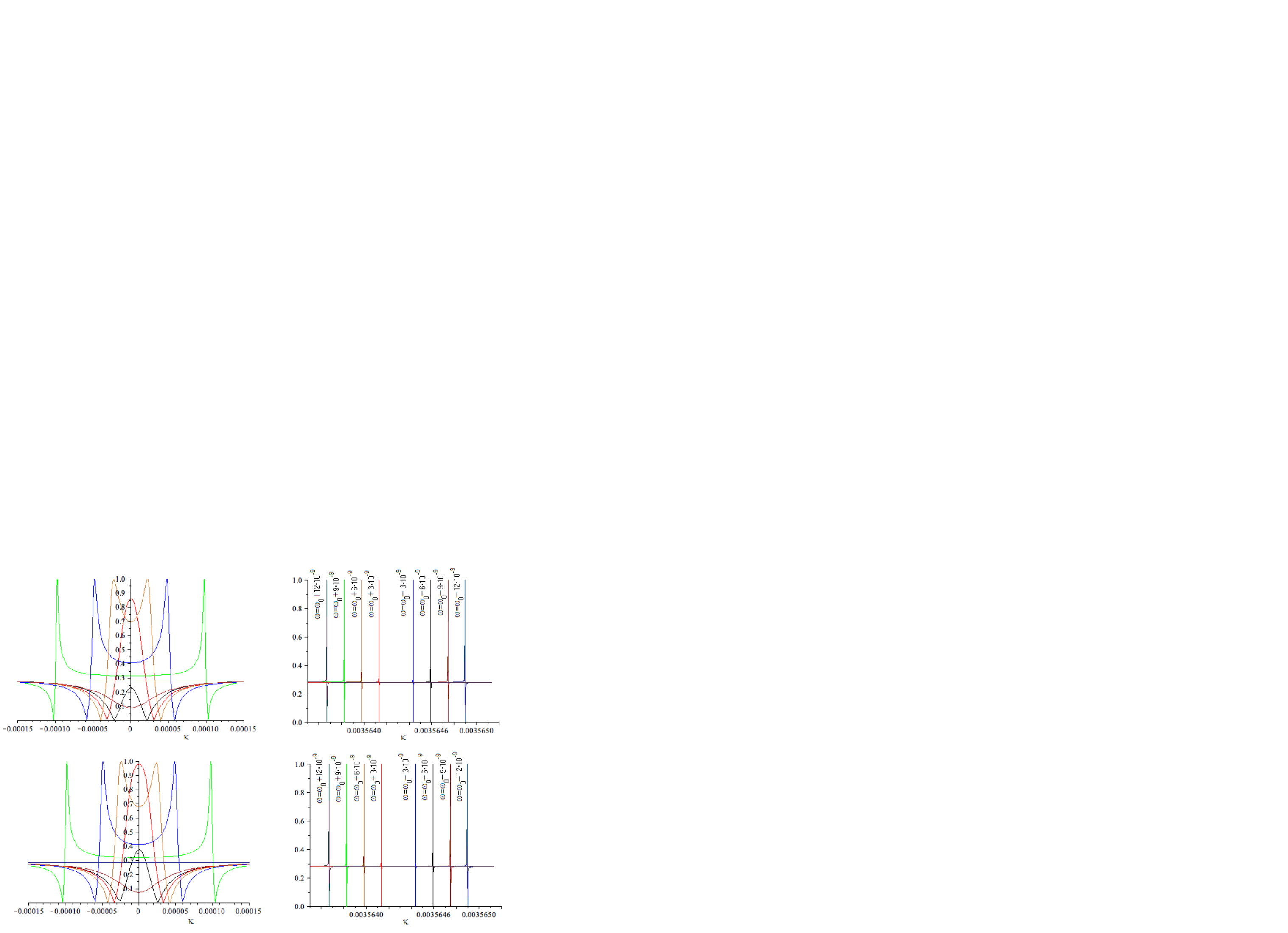}
\caption{\footnotesize{After bifurcation: The transmission coefficient for $M_0=2$, $M_1=1$, $k_1=k_2=1$, $\gamma_0=\gamma_0^*-0.0001=1.029533513$, $\gamma_1=7$ near $\kappa=0$ and near the parameters of the guided mode, $\kappa_0=0.003564296929$, $\omega_0=0.9778903229$. Upper: Direct calculation of the transmission coefficient. Lower: The asymptotic formula for $|b|/|\ell|$.}}
 \label{bifur_along}
\end{figure}

Figs.~\ref{bifur_itself} and \ref{bifur_along} show the transmission coefficient at and after the bifurcation, both by direct calculation as well as using the above expansions in the expression $|b|/|\ell| = |b|/\sqrt{|a|^2+|b|^2}$ with appropriate choices of coefficients up to quadratic order.



\subsection{Resonant Enhancement}

The transmission anomalies that we have analyzed are accompanied by the perhaps more elementary phenomenon of resonant enhancement of the amplitude of the field in the waveguide.
This enhancement will be manifest in the resonant component of the field $\overrightarrow{X}$, which is the first term of equation \eqref{Xsolution},
\begin{equation}\label{Xsolution2}
\vec{X}=c\alpha Ce_1+c\ell C(0, \mathbb{\widetilde{B}}^{-1}F_2).
\end{equation}
Thus a good measurement of amplitude enhancement is the ratio $|\alpha/\ell|$. Let $\alpha$ in the vicinity of $(\kappa_0,\omega_0)$ have the expansion
\begin{equation}
\label{expensionalpha}
\alpha = \beta_0+\beta_1\widetilde{\kappa}+\beta_2\widetilde{\omega}+\cdots
\end{equation}
{\theorem In the expansion $(\ref{expensionalpha})$, $\beta_0=0$.}
\vspace{1ex}

\noindent {\emph{Proof:}} By Theorem~\ref{solution}, at the pair $(\kappa_0,\omega_0)$ there is a solution $\overrightarrow{X}$ to the scattering problem $\mathbb{B}\vec{X}=\vec{F}$,
with $\overrightarrow{F}=C(\alpha,F_2)$.  Using $\ell(\kappa_0,\omega_0)=0$, this equation becomes
\begin{equation}
\label{alphanol}
\left(
\begin{matrix}
0 & 0\\
0 & \mathbb{\widetilde{B}}
\end{matrix}
\right)
C^{-1}\overrightarrow{X}
=
\mathbb{J}C^{-1}\overrightarrow{X}=
C^{-1}\mathbb{B}\overrightarrow{X}
=
C^{-1}\overrightarrow{F}
=
\left(
\begin{matrix}
\alpha\\
F_2
\end{matrix}
\right).
\end{equation}
It follows that $\alpha(\kappa_0,\omega_0)=0$. 
\QED
 
\medskip
Using the forms for $\alpha$ and $(\ref{forms})$ for $\ell$ , we obtain
\begin{equation}
\label{ratio}
\frac{\alpha}{\ell} = \frac{\beta_1\widetilde{\kappa}+\beta_2\widetilde{\omega}+\cdots}{(\widetilde{\omega}+\ell_1\widetilde{\kappa}+\ell_2\widetilde{\kappa}^2+\cdots)}\left(\frac{1}{e^{i\rho_1}}+\cdots \right)
\end{equation}

\begin{figure}
\centering 
\includegraphics[width=13.0cm,height=5.1cm]{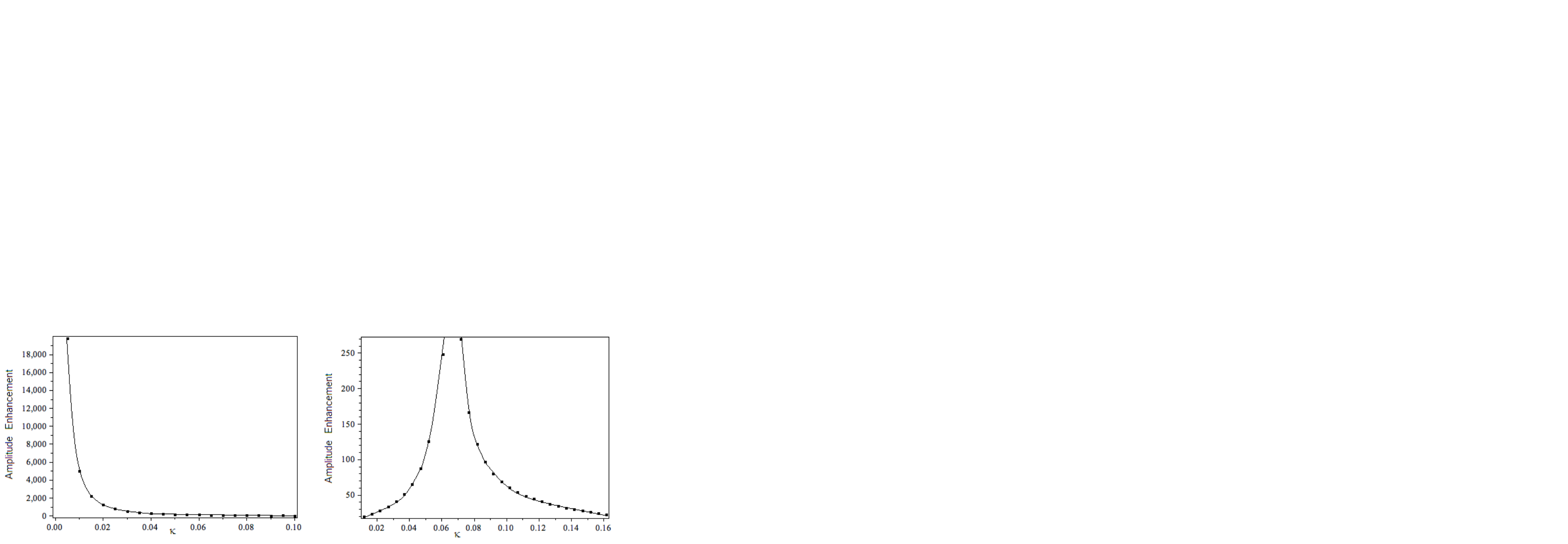}
\caption{\footnotesize{The dots represent numerically calculated magnitudes of the field in the waveguide produced by an incident plane wave of amplitude 1 at various values of $\kappa$ and at $\omega=\omega_0-\ell_{1}\wk-\re(\ell_{2})\wk^{2}$ for two cases of period $N=2$.   Left: $\kappa_0^*=0$, with $M_0=2$, $M_1=1$, $k_0=k_1=1$, $\gamma_0=1.029633513$, $\gamma_1=7$. Right: $\kappa_0=0.0616$, with $M_0=2$, $M_1=1$, $k_0=k_1=1$, $\gamma_0=1$, $\gamma_1=7$.}}
\label{ResonEnh}
\end{figure}
%
When $\wk$ is small, the magnitude of the denominator in $\alpha/\ell$ is minimized to order $\bigo(\wk^2)$ when $\wo+\ell_1\wk+\re(\ell_2)\wk^2$. To see the response to an incident plane wave at this optimal frequency, put
\begin{equation}
\wo=-\ell_1\wk-\re(\ell_2)\wk^2, \;\text { or }\; \omega=\omega_0-\ell_1\wk-\re(\ell_2)\wk^2,
\end{equation}
and obtain for the amplitude enhancement $\mathcal{A}$
\begin{equation}
\mathcal{A}=\Bigl| \frac{\alpha}{\ell} \Bigr| = \frac{1}{ \wk} \Bigl| \frac{\beta_{1}-\ell_1 \beta_{2} -\beta_2 \re(\ell_2)\wk+\cdots}{i \im(\ell_2)+\cdots}\Bigr|
\end{equation}
so that $\mathcal{A}$ has an asymptotic expansion of the form
\begin{equation}
\label{asymptotic}
\mathcal{A} \sim \frac{d_1}{\wk}+d_2+\cdots  \quad (\wo=-\ell_{1}\wk-\re(\ell_{2})\wk^{2}, \quad \wk \to 0)
\end{equation}
Fig.~\ref{ResonEnh} shows numerical calculations that confirm the $1/\wk$ behavior of the amplitude of the waveguide at $\wo=-\ell_{1}\wk-\re(\ell_{2})\wk^{2}$ for $N=2$.  The magnitude of the field is calculated using the expression $\sqrt{|c_0|^2+|c_1|^2}$.


\section{Appendix:  Difference Operators}

We use the following notation:
\begin{eqnarray*}
  && v = \{v_n\}, \quad w = \{w_n\}, \quad vw = \{v_nw_n\}, \\
  && (v^+)_n = v_{n+1}, \quad (v^-)_n = v_{n-1}, \\
  && (\partial_x v)_n = (v_x)_n = v_{n+1}-v_n, \quad (\partial_{\bar x} v)_n = (v_{\bar x})_n = v_n - v_{n-1}.
\end{eqnarray*}
One can compute the discrete product rule and the fundamental theorem as well as a summation-by-parts formula:
\begin{eqnarray}
  && (vw)_x = v_xw^+ + vw_x = v_x w + v^+ w_x, \label{discrete1a}\\
  && (vw)_{\bar x} = v_{\bar x}w^- + vw_{\bar x} = v_{\bar x}w + v^- w_{\bar x}, \label{discrete1b}\\
  && v_{m_2}-v_{m_1} = \sum_{m=m_1+1}^{m_2} (v_{\bar x})_m = \sum_{m=m_1}^{m_2-1}(v_x)_m, \label{discrete1c}\\
  && \sum_{m=m_1+1}^{m_2} (\partial_{\bar x}v)_mw_m 
       + \sum_{m=m_1+1}^{m_2}v^-_m\partial_{\bar x}w_m = v_{m_2}w_{m_2} - v_{m_1}w_{m_1}.
       \label{discrete1d}
\end{eqnarray}
%


For the two-dimensional discrete calculus, we use the notation
\begin{eqnarray*}
 && \mathbf{F} = \{\mathbf{F}_{mn}\} = \{(F^1_{mn},F^2_{mn})\}, \\
 && (\mathbf{F}^-)_{mn} = (F^1_{m-1,n},F^2_{m,n-1}),\\
 && \nabla_{\!-} = (\partial_{\bar x},\partial_{\bar y}), \; \nabla_{\!+} = (\partial_{x},\partial_{y}).
\end{eqnarray*}
One has the discrete product rule
\begin{equation}\label{productrule}
  \nabla_{\!-}\cdot (v\mathbf{F}) = v\nabla_{\!-}\cdot\mathbf{F} + \nabla_{\!-} v\cdot \mathbf{F}^-.
\end{equation}
The discrete divergence theorem for a rectangular region $[m_1,m_2]\!\times\![n_1,n_2]$ is
\begin{equation}
\label{thirty-fourth}
\sum\limits_{n=n_1 +1}^{n_2}\sum\limits_{m=m_1 +1}^{m_2}(\nabla_{\!-}\!\cdot\!\mathbf{F})_{mn}=\sum\limits_{n= n_1 +1}^{n_2} (F^{1}_{m_{2}  n}- F^{1}_{m_{1} n})+ \sum\limits_{m= m_1 +1}^{m_2}(F^{2}_{m n_{2}}-F^{2}_{m n_{1}}).
\end{equation}
Now put $\mathbf{F}=v\nabla_{\!+}u$ in \eqref{thirty-fourth} and expand $\nabla_{\!-}\cdot(v\nabla_{\!+}u)$ using \eqref{productrule} to obtain the two-dimensional summation-by-parts identity
\begin{equation}
\label{summation}
\begin{array}{rcl}
\sum\limits_{n=n_1 +1}^{n_2} \sum\limits_{m=m_1 +1}^{m_2}(v \Delta u)_{mn}&=&\sum\limits_{n=n_1 +1}^{n_2}((vu_{x})_{m_2 n}-(vu_{x})_{m_1 n})\\
& & +\sum\limits_{m=m_1 +1}^{m_2}((vu_{y})_{m n_2}-(vu_{y})_{m n_1}) \\
& & -\sum\limits_{n=n_1 +1}^{n_2} \sum\limits_{m=m_1 +1}^{m_2} (\nabla_{\!-} v \cdot \nabla_{\!-}u)_{mn}.
\end{array}
\end{equation}
Here, one uses the identities $\nabla_{\!-}\cdot\nabla_{\!+}u = \Delta u$ and
$(\nabla_{\!+}u)^- = \nabla_{\!-}u$.

\vspace{2ex}
\centerline{\bfseries\Large Acknowledgment}

\vspace{1ex}
\noindent
Both authors are grateful for the support of NSF grants DMS-0505833 and DMS-0807325.  N.~Ptitsyna thanks the
Louisiana State Board of Regents for support under the Student Travel Grant LEQSF(2005-2007)-ENH-TR-21.

\bibliography{PtitsynaShipman}

\end{document}